\documentclass{aa}

\usepackage{graphicx}
\usepackage{txfonts}
\usepackage{hyperref}
\usepackage{multirow}
\usepackage{textcomp}
\begin{document}

\title{Simulating the exoplanet yield of a space-based mid-infrared interferometer based on \emph{Kepler} statistics}

\subtitle{}

\author{Jens Kammerer \& Sascha P. Quanz\thanks{National Center of Competence in Research "PlanetS" (\url{http://nccr-planets.ch})}}

\institute{ETH Zurich, Institute for Particle Physics and Astrophysics, Wolfgang-Pauli-Strasse 27, 8093 Zurich, Switzerland \\
\email{sascha.quanz@phys.ethz.ch}}
%

\date{Received --- ; accepted --- }

\abstract{}{}{}{}{}

\abstract
{}
{We predict the exoplanet yield of a space-based mid-infrared nulling interferometer using Monte Carlo simulations. We quantify the number and properties of detectable exoplanets and identify those target stars that have the highest or most complete detection rate. We investigate how changes in the underlying technical assumptions and uncertainties in the underlying planet population impact the scientific return.}
{We simulated $2'000$ exoplanetary systems, based on planet occurrence statistics from \emph{Kepler} with randomly orientated orbits and uniformly distributed albedos around each of $326$ nearby ($d < 20~\text{pc}$) stars. Assuming thermal equilibrium and blackbody emission, together with the limiting spatial resolution and sensitivity of our simulated instrument in the three specific bands $5.6$, $10.0$, and $15.0~\text{\textmu m}$, we quantified the number of detectable exoplanets as a function of their radii and equilibrium temperatures.}
{Approximately $\sim315_{-77}^{+113}$ exoplanets, with radii $0.5~R_\text{Earth} \leq R_\text{p} \leq 6~R_\text{Earth}$, were detected in at least one band and half were detected in all three bands during $\sim0.52$ years of mission time assuming throughputs $3.5$ times worse than those for the James Webb Space Telescope and $\sim40\%$ overheads. Accounting for stellar leakage and (unknown) exozodiacal light, the discovery phase of the mission very likely requires $2$ -- $3$ years in total. The uncertainties in planet yield are dominated by uncertainties in the underlying planet population, but the distribution of the Bond albedos also has a significant impact. Roughly 50\% of the detected planets orbit M stars, which also have the highest planet yield per star; the other 50\% orbit FGK stars, which show  a higher completeness in the detectability. Roughly $85$ planets could be habitable ($0.5~R_\text{Earth} \leq R_\text{p} \leq 1.75~R_\text{Earth}$ and $200~\text{K} \leq T_\text{eq} \leq 450~\text{K}$) and are prime targets for spectroscopic observations in a second mission phase. Comparing these results to those of a large optical/near-infrared telescope, we find that a mid-infrared interferometer would detect more planets and the number of planets depends less strongly on the wavelength.}
{An optimized space-based nulling interferometer operating in the mid-infrared would deliver an unprecedented dataset for the characterization of (small) nearby exoplanets including dozens of potentially habitable worlds.}


\titlerunning{Exoplanet yield of a space-based mid-infrared interferometer}

\maketitle


\section{Introduction}

The detection and characterization of extrasolar planets is one of the most dynamic research fields in modern astrophysics. To date more than $5'200$ exoplanets and exoplanet candidates\footnote{\url{http://exoplanets.org} (as of July 17, 2017)} have been identified by various detection techniques and instruments on ground and in space. In particular the NASA \emph{Kepler} mission \citep{borucki2010} revolutionized the field by providing, for the first time, robust statistics for the occurrence rate of exoplanets as a function of their size, orbital period, and stellar host type, revealing that small-sized planets with radii $< 2~R_\text{Earth}$ are ubiquitous \citep[e.g.,][]{howard2012, dressing2013, fressin2013, mulders2015, dressing2015, burke2015, gaidos2016}. Most of these studies make use of various pipeline completeness models to estimate the distribution of exoplanets in the radius-orbital period space depending on the spectral type of the host star and give their results in either continuous probability density functions or binned occurrence rates. While \emph{Kepler} was an unprecedented success in constraining the occurrence rate of exoplanets even for planets with radii down to $1~R_\text{Earth}$, most of the \emph{Kepler} systems are too distant for in-depth follow-up observations for atmospheric characterization. With the upcoming \emph{TESS} mission from NASA and then later the ESA \emph{PLATO} mission, the number of transiting planets detected around bright, nearby stars will significantly increase. Furthermore, a few transiting systems in the immediate solar neighborhood have already been identified, some of which are excellent targets for follow-up observations with the James Webb Space Telescope (\emph{JWST}) even for the search of biomarkers \citep[e.g.,][]{bertathompson2015, gillon2017a, gillon2017b}. However, it is important to keep in mind that transit observations will always only probe a small subset of the exoplanet population as the transit probability is inversely proportional to the star-planet separation. So, overall, it is unclear how many small-sized planets will be uncovered from transit surveys that are suitable for in-depth follow-up observations via transit or secondary eclipse spectroscopy and, at the same time, cover a broad range of host star spectral types and orbital separations.

In this context, a few studies investigated to what extent the next generation of extremely large ground-based telescopes (ELTs) can be used to detect and characterize small-sized exoplanets \citep[e.g.,][]{snellen2015}. \citet{crossfield2013} and \citet{quanz2015} used statistical results from the \emph{Kepler} mission to model exoplanetary systems around nearby stars with the goal of estimating the number and characteristics of exoplanets directly detectable by different ground-based high-contrast imaging instruments such as the mid-infrared E-ELT imager and spectrograph (E-ELT/METIS). Unlike the transit method, direct imaging, which aims at spatially separating the flux received from an exoplanet and the flux received from its host star, is primarily sensitive to widely separated planets as the contrast achievable by the instrument increases with increasing angular separation. It turns out that the spatial resolution and the sensitivity of the next generation of ground-based telescopes will enable the detection of a few small-sized planets around the very nearest stars maybe even in the habitable zone \citep{quanz2015}, but the characterization of a sizeable sample of small-sized exoplanets very likely remains out of reach even in the era of the ELTs.

Already more than a decade ago mid-infrared (MIR) space-based interferometers, providing unprecedented spatial resolution and sensitivity for the search of exoplanets, were proposed to both NASA \citep[Terrestrial Planet Finder, \emph{TPF};][]{lawson2001} and ESA \citep[\emph{Darwin};][]{leger2007}. Unfortunately, neither of the two missions were implemented; among several technical challenges, one scientific reason for the lack of implementation were uncertainties in their expected exoplanet yield. \emph{Kepler} released its first data only in 2010 \citep{borucki2010} and also ground-based radial velocity surveys had hardly begun to uncover exoplanets in the super-Earth mass regime at that time. In consequence, the occurrence rate of small, terrestrial planets was largely unconstrained when \emph{TPF} and \emph{Darwin} were discarded. More recently several papers have investigated the capabilities of space-based direct imaging missions to detect and characterize exoplanets focusing on Earth twins in environments appropriate for extrasolar life \citep[e.g.,][]{stark2014, brown2015, leger2015, stark2015}. These studies indeed made use of a statistical framework of Earth twins inferred from \emph{Kepler} data. However, all of these studies focused solely on planets that are distinctly equal to our own Earth. This is reflected in their choice of physical planet parameters, which they assign to their simulated exoplanet sample that is necessary to estimate the scientific yield of their hypothetical space observatories. It is beyond debate that the search for extrasolar life is one main goal of exoplanet research, but to understand the complicated processes involved in the formation and evolution of planets, a diverse and comprehensive sample of exoplanets is needed. Furthermore, the occurrence of Earth-like planets in the habitable zone of their parent stars is still affected by high uncertainties, varying from a factor of $\sim2$ around M-type stars \citep[e.g.,][]{dressing2015} to a factor of $\sim10$ around G- and K-type stars \citep[e.g.,][]{burke2015}.

Here, for the first time to our knowledge, we quantify the general scientific yield of a large space-based MIR interferometer by combining the technical specifications of the \emph{Darwin} mission, as proposed a decade ago, with the most recent planet occurrence statistics from the \emph{Kepler} mission. Using Monte Carlo simulations and a stellar sample of $326$ nearby stars within $20~\text{pc}$, we make predictions for the detection rates of exoplanets in different bands between $5.6$ and $15~\text{\textmu m}$ and over a large range of planet radii ($0.5$ -- $6~R_\text{Earth}$) and orbital periods ($0.5$ -- $418~\text{d}$). Our analysis allows us to identify the best target stars and to estimate what percentage of exoplanets might also be detectable with future, high-precision radial velocity or space-based optical/near-infrared (NIR) instruments because fully characterizing a planet, and in particular assessing its habitability, will require data from multiple techniques and instruments over a wavelength range that is as broad as possible. We also quantify what scientific gain/loss, in terms of planet detections, one can expect depending on the achievable spatial resolution and sensitivity of the space interferometer. As such, this paper aims at providing a \emph{scientific} basis for reinitiating and inspiring the discussion for the need for such a mission in the long term; we do not discuss the technical readiness or related challenges.


\section{Methods}

In our Monte Carlo simulations we model $2'000$ exoplanetary systems around each of $326$ nearby stars within $20~\text{pc}$ based on planet occurrence statistics from \emph{Kepler}. All astrophysical and instrumental parameters for our baseline scenario can be found in Table~\ref{baseline} and in the following subsections.

\begin{table}[!h]
\caption{\label{baseline}Astrophysical and instrumental parameters for our baseline scenario.}
\centering
\begin{tabular}{lll}
\hline\hline\noalign{\smallskip}
Parameter              & Value                    & Description \\
\noalign{\smallskip}\hline\noalign{\smallskip}
$\cos i$               & $[-1,~1)$                & Cosine of inclination \\
$\Omega$               & $[0,~2\pi)$              & Longitude of ascending node \\
$\omega$               & $[0,~2\pi)$              & Argument of periapsis \\
$\vartheta$            & $[0,~2\pi)$              & True anomaly \\
$e$                    & $0$                      & Eccentricity\tablefootmark{a} \\
$A_\text{B}$           & $[0,~0.8)$               & Bond albedo\tablefootmark{b} \\
$A_\text{g}$           & $[0,~0.1)$               & Geometric albedo\tablefootmark{b} \\
\noalign{\smallskip}\hline\noalign{\smallskip}
IWA                    & $5.0~\text{mas}$         & IWA @ $\lambda_\text{eff} = 10~\text{\textmu m}$\tablefootmark{c, d} \\
OWA                    & $1''$                    & OWA @ $\lambda_\text{eff} = 10~\text{\textmu m}$\tablefootmark{c, e} \\
$F_\text{lim,~F560W}$  & $0.16~\text{\textmu Jy}$ & Sensitivity limit ($\lambda_\text{eff} = 5.6~\text{\textmu m}$)\tablefootmark{f} \\
$F_\text{lim,~F1000W}$ & $0.54~\text{\textmu Jy}$ & Sensitivity limit ($\lambda_\text{eff} = 10.0~\text{\textmu m}$)\tablefootmark{f} \\
$F_\text{lim,~F1500W}$ & $1.39~\text{\textmu Jy}$ & Sensitivity limit ($\lambda_\text{eff} = 15.0~\text{\textmu m}$)\tablefootmark{f} \\
\noalign{\smallskip}\hline
\end{tabular}
\tablefoot{Values are distributed uniformly in the given ranges. \\
\tablefoottext{a}{see discussion in Section~\ref{eccentricity}} \\
\tablefoottext{b}{see discussion in Section~\ref{albedos}} \\
\tablefoottext{c}{from \citet{leger2007}} \\
\tablefoottext{d}{IWA = inner working angle; scales with $\lambda_\text{eff}$} \\
\tablefoottext{e}{OWA = outer working angle; scales with $\lambda_\text{eff}$} \\
\tablefoottext{f}{$10\sigma$ detection limit in $35'000~\text{s}$; modified from \citet{glasse2015} who assumed $10'000~\text{s}$ to achieve these limits with \emph{JWST}/MIRI}
}
\end{table}


\begin{figure*}[h!]
\centering
\includegraphics[width=9cm]{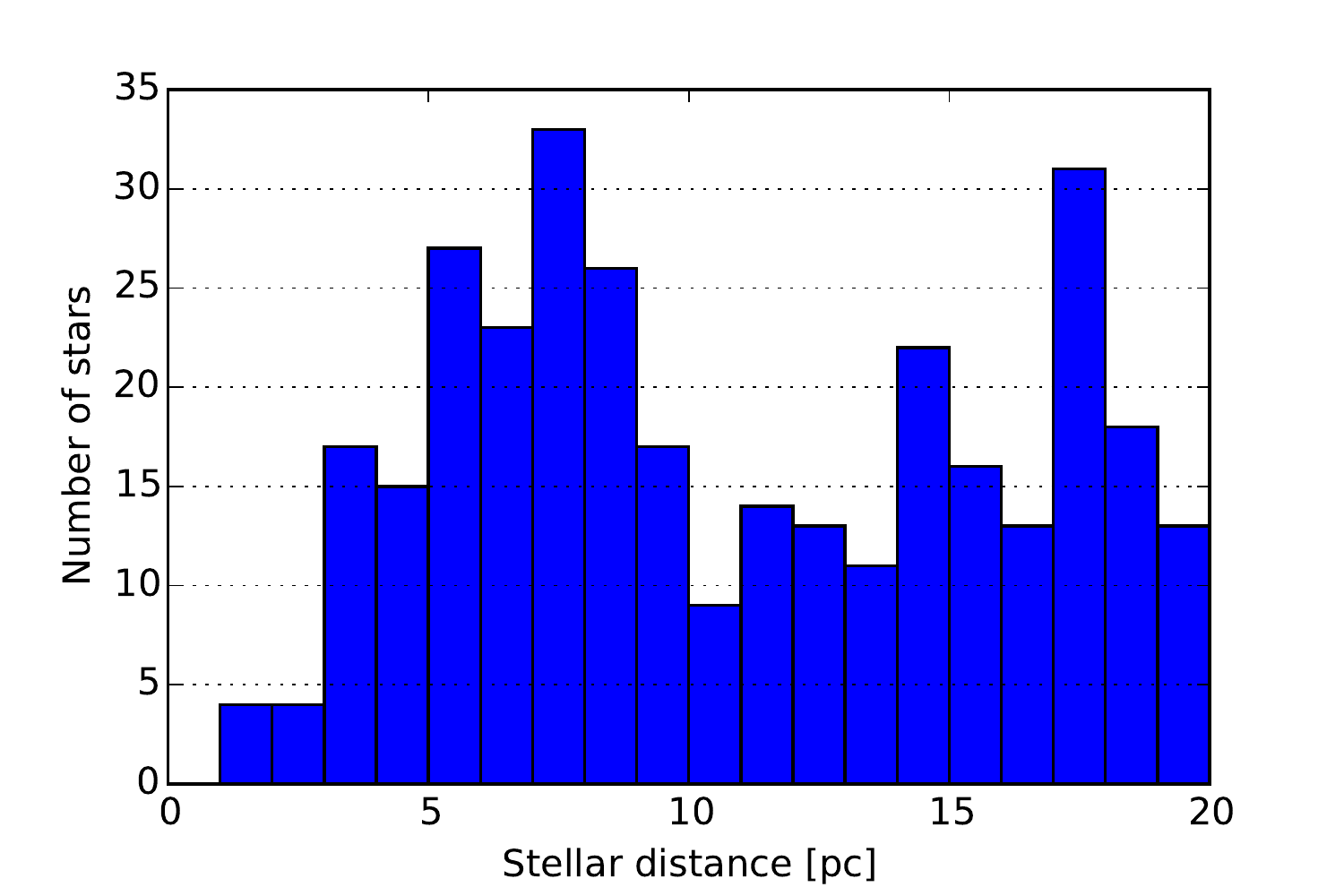}
\includegraphics[width=9cm]{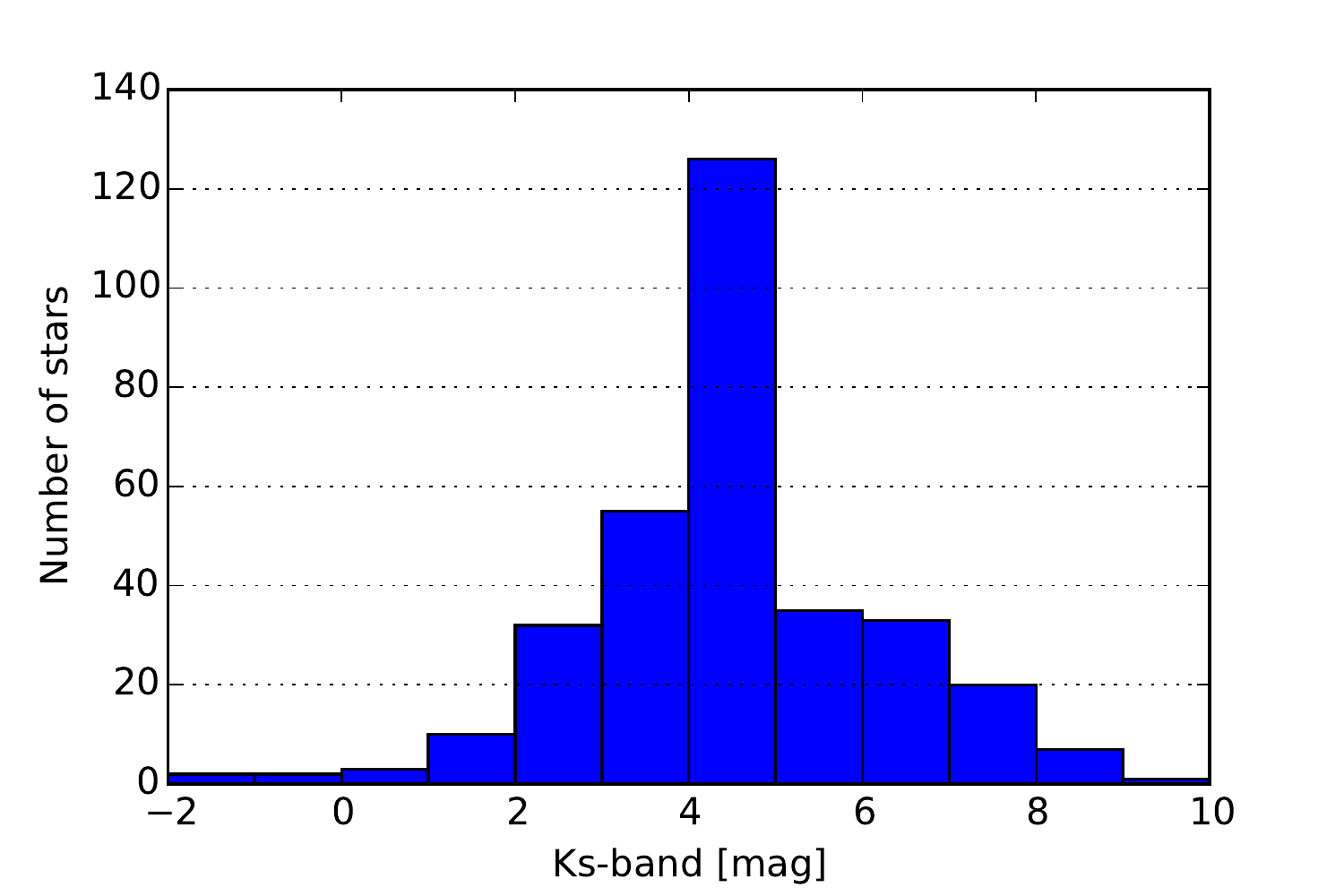}
\caption{Left: Distribution of the stellar distances in parsec including all stars from our star catalog. Right: Distribution of the measured Ks-band magnitudes from \emph{2MASS} for the same stars.}
\label{newfig_01}
\end{figure*}


\subsection{Stellar sample}

We adopted the star catalog from the E-ELT/METIS study presented in \citet{quanz2015} containing $328$ stars of spectral types A, F, G, K and M out to a distance of $21.4~\text{pc}$, but we removed two stars with parallaxes $< 50~\text{mas}$ to obtain a stellar sample limited to a distance of $20~\text{pc}$. The basis for this catalog was originally established by \citet{kirkpatrick2012} and contained objects within $8~\text{pc}$. Then white dwarfs and stars with spectral types later than M7 and binaries with apparent angular separations $< 5''$ were removed by \citet{crossfield2013}. Stars with fainter companions were retained. The star catalog was finally appended by \citet{quanz2015} with all dwarf stars with K-band magnitudes $< 7^\text{mag}$ out to a distance of $10~\text{pc}$ and $< 5^\text{mag}$ out to a distance of $20~\text{pc}$ from the SIMBAD Astronomical Database\footnote{\url{http://simbad.u-strasbg.fr/simbad/}}. Close binaries were again removed and empirical relations were used to convert magnitudes and spectral types into stellar parameters. The final catalog contains stellar properties (e.g., parallax, spectral type, radius, effective temperature, and mass) for $326$ stars ($8$ A stars, $54$ F stars, $72$ G stars, $71$ K stars, and $121$ M stars) distributed over the whole sky. Figure~\ref{newfig_01} shows, for illustrative purposes, histograms of the stellar distances and the Ks-band magnitudes as measured from \emph{2MASS} \citep{skrutskie2006}.


\subsection{Planet population}
\label{planet_population}

To create a random planet population we used planet occurrence statistics from \citet{fressin2013}, \citet{dressing2015}, and \citet{burke2015}. \citet{fressin2013} covered the broadest range of planet radii from $0.8$ to $22~R_\text{Earth}$ and orbital periods from $0.8$ to $418~\text{d}$ corresponding to $0.017$ to $1.094~\text{au}$ around a Sun-like star. As \citet{fressin2013} did not find a significant dependence of the planet occurrence on the spectral type we applied these statistics to all AFGKM stars in our star catalog.

As we are especially interested in terrestrial exoplanets, whenever possibly we decided to use the planet occurrence statistics from \citet{dressing2015} and \citet{burke2015} instead of those from \citet{fressin2013} for small exoplanets. Compared to \citet{fressin2013}, who only used Q1 -- Q6 data to identify \emph{Kepler} planet candidates, \citet{dressing2015} and \citet{burke2015} considered Q0 -- Q17 and Q1 -- Q16 data, respectively. Therefore, we expected their statistics to be more complete. Furthermore, they concentrated on certain spectral types, which allowed us to take into account a spectral type dependence of the planet occurrence rates. This patchwork-like approach is summarized in Table~\ref{ragrug}.

\begin{table}[!h]
\caption[]{\label{ragrug}Applied planet occurrence statistics with the planet radius $R_\text{p}$ and orbital period $P_\text{orb}$ ranges for each spectral type appearing in our star catalog.}
\centering
\begin{tabular}{llll}
\hline\hline\noalign{\smallskip}
Spec. type         & Statistics & $R_\text{p}$ [$R_\text{Earth}$] & $P_\text{orb}$ [d] \\
\noalign{\smallskip}\hline\noalign{\smallskip}
A                  & Fressin    & 0.8 -- 22                       & 0.8 -- 418 \\
\noalign{\smallskip}\hline\noalign{\smallskip}
F                  & Fressin    & 0.8 -- 22                       & 0.8 -- 418 \\
\noalign{\smallskip}\hline\noalign{\smallskip}
\multirow{4}{*}{G} & Burke      & 0.75 -- 2.5                     & 50 -- 300 \\
                   & Fressin    & 0.8 -- 22                       & 0.8 -- 50 \\
                   & Fressin    & 0.8 -- 22                       & 300 -- 418 \\
                   & Fressin    & 2.5 -- 22                       & 50 -- 300 \\
\noalign{\smallskip}\hline\noalign{\smallskip}
\multirow{4}{*}{K} & Burke      & 0.75 -- 2.5                     & 50 -- 300 \\
                   & Fressin    & 0.8 -- 22                       & 0.8 -- 50 \\
                   & Fressin    & 0.8 -- 22                       & 300 -- 418 \\
                   & Fressin    & 2.5 -- 22                       & 50 -- 300 \\
\noalign{\smallskip}\hline\noalign{\smallskip}
\multirow{3}{*}{M} & Dressing   & 0.5 -- 4                        & 0.5 -- 200 \\
                   & Fressin    & 0.8 -- 22                       & 200 -- 418 \\
                   & Fressin    & 4 -- 22                         & 0.8 -- 200 \\
\noalign{\smallskip}
\hline
\end{tabular}
\tablefoot{Planet radii and orbital periods are not simulated in the same range throughout all spectral types.
}
\end{table}

The planet occurrence statistics from \citet{fressin2013} and \citet{dressing2015} were binned in planet radius and orbital period space. We applied a Poisson distribution for the number of planets per star and drew planet radii uniformly in linear space inside a single bin \citep[e.g.][]{crossfield2013} and orbital periods uniformly in logarithmic space inside a single bin in agreement with \citet{petigura2013} and \citet{silburt2015}. We conservatively assumed that the planet occurrence rate is zero inside bins for which no statistics are provided. If the $1\sigma$ upper limits are given instead of the statistical occurrence of exoplanets $r$ we proceeded as follows: We calculated the missing $r$ using the cumulative number of planets per star versus orbital period from Table~5 of \citet{dressing2015}. For $3.5~R_\text{Earth} \leq R_\text{p} \leq 4~R_\text{Earth}$, where $r$ is missing for two orbital period bins we assumed that the statistical occurrence of exoplanets is the same for the $3.5~R_\text{Earth} \leq R_\text{p} \leq 4~R_\text{Earth}$ and $0.5~\text{d} \leq P_\text{orb} \leq 1.7~\text{d}$ bin and the $3~R_\text{Earth} \leq R_\text{p} \leq 3.5~R_\text{Earth}$ and $0.5~\text{d} \leq P_\text{orb} \leq 1.7~\text{d}$ bin (including its errors). We further assumed the upper error to be the difference between our calculated $r$ and the given $1\sigma$ upper limit. If the obtained upper error is smaller than $r$ we set the lower error equal to the upper error. However, if the obtained upper error is larger than $r$ we set the lower error equal to $r$ itself to avoid negative planet occurrence rates.

\citet{burke2015} provided their planet occurrence statistics in terms of a continuous probability density function,
\begin{equation}
\label{eq01}
\frac{d^2f}{dR_\text{p}dP_\text{orb}} = F_0C_\text{n} \cdot \begin{cases} \left( \frac{R_\text{p}}{R_0} \right)^{\alpha_1}\left( \frac{P_\text{orb}}{P_0} \right)^\beta &\quad R_\text{p} < R_\text{brk}, \\ \left( \frac{R_\text{p}}{R_0} \right)^{\alpha_2}\left( \frac{R_\text{brk}}{R_0} \right)^{\alpha_1-\alpha_2}\left( \frac{P_\text{orb}}{P_0} \right)^\beta &\quad R_\text{p} \geq R_\text{brk}, \end{cases}
\end{equation}
where we also applied a Poisson distribution with mean value $F_0$ for the number of planets per star since
\begin{equation}
\label{eq02}
\int_{R_\text{p,~min}}^{R_\text{p,~max}}\int_{P_\text{orb,~min}}^{P_\text{orb,~max}}\frac{f}{F_0}dR_\text{p}dP_\text{orb} = 1,
\end{equation}
and thus $F_0$ represents the average number of planets per star. We separated the continuous probability density function into a planet radius and an orbital period part and calculated their inverse cumulative distribution functions (normalized to one) into which a uniformly distributed random variable between 0 and 1 can be inserted yielding a planet radius and an orbital period distribution according to Equation~\ref{eq01}.

We note that orbital periods $\leq 418~\text{d}$ are too short to probe the habitable zones around A- and F-type stars. Furthermore, we disregard the conservative habitable zone around G-type stars \citep{kopparapu2013}, Earths ($0.8$ -- $1.25~R_\text{Earth}$), and super-Earths ($1.25$ -- $2~R_\text{Earth}$), since \citet{fressin2013} provided empty bins above $P_\text{orb} = 145~\text{d}$ and the data from \citet{burke2015} extended to $P_\text{orb} = 300~\text{d}$ only. However, as mentioned in the Introduction, we do not solely focus on exoplanets in the habitable zone here and feel uncomfortable with extrapolating the orbital period distribution to longer periods for our baseline scenario. We come back to the question of potentially habitable exoplanets in Section~\ref{exo_earths}, however.

We assigned a randomly distributed orbit and position on this orbit and a randomly drawn Bond and geometric albedo to each generated exoplanet. We distributed the Bond albedos $A_\text{B}$ uniformly in $[0,~0.8)$ considering the Bond albedos of the solar system planets, which range from $0.068$ (Mercury) to $0.77$ (Venus)\footnote{\url{http://nssdc.gsfc.nasa.gov/planetary/factsheet/}} and the geometric albedos $A_\text{g}$ uniformly in $[0,~0.1)$, because many gases that dominate the atmospheres of our solar system planets absorb strongly in the infrared \citep[e.g., $\text{H}_2\text{O}$, $\text{CO}_2$, $\text{O}_3$, $\text{CH}_4$, $\text{N}_2\text{O}$;][]{green1964}. As we discuss further below the geometric albedos have no significant influence on our results since the percentage of exoplanets that can be detected only due to their reflected host star light is negligible longward of $\sim3~\text{\textmu m}$. We then calculated the apparent angular separation between exoplanet and host star and the effective temperature of the exoplanet $T_\text{eff,~p}$, which we assumed to be equal to its equilibrium temperature,
\begin{equation}
\label{eq03}
T_\text{eff,~p} = T_\text{eq,~p} = \left[ \frac{R_*^2(1-A_\text{B})}{4r_\text{p}^2} \right]^\frac{1}{4}T_{\text{eff},~*},
\end{equation}
where $R_*$ is the radius of the host star, $r_\text{p}$ is the physical separation between exoplanet and host star, and $T_{\text{eff},~*}$ is the effective temperature of the host star. We estimated the emitted thermal flux $F_\text{therm,~p}$ of the exoplanet, assuming it emits similar to a blackbody with temperature $T_\text{eff,~p}$. Finally, we computed the reflected host star flux from the exoplanet according to
\begin{equation}
\label{eq04}
F_\text{refl,~p} = A_\text{g}f(\alpha)\frac{R_\text{p}^2}{d^2}F_{\text{inc},~*},
\end{equation}
where $f(\alpha)$ is the exoplanet's phase curve, $d$ is the distance to Earth, and $F_{\text{inc},~*}$ is the host star flux, which is incident on the exoplanet, assuming Lambertian scattering \citep[cf., e.g.,][]{seager2010} and that the host stars are spherical blackbodies as well.


\subsection{Instrument properties}

Unlike other studies \citep[e.g.,][]{crossfield2013, stark2014, brown2015, stark2015}, which have investigated the exoplanet yield for a broad range of instrument parameters, we focused on a specific instrument configuration that was proposed to ESA in 2007. We took the technical specifications (see bottom half of Table~\ref{baseline}) of this space-based nulling interferometer from the proposal published by \citet{cockell2009}. This proposal suggested an instrument consisting of four formation-flying mirrors that focus their light onto a fifth beam combiner spacecraft (BCS), where detectors and communication devices are located. The spacecraft would be arranged in an X-like architecture with the four mirrors flying in a single plane at the tips of the X and effectively forming part of a large synthetic paraboloid. The BCS would fly $\sim1.2~\text{km}$ above this plane in the center of the X. The whole setup could be launched to the Earth-Sun L2 point in a single Ariane 5 rocket and would be able to observe an annular region between $46^\circ$ and $83^\circ$ from antisolar direction and thus reach complete sky coverage ($> 99\%$) throughout the whole year.

Besides an imaging baseline $B_\text{I}$ of up to $500~\text{m}$, a nulling baseline $B_\text{N}$ from $7$ to $168~\text{m,}$ resulting in a best possible IWA (inner working angle) of $\lambda_\text{eff}/(4B_\text{N}) \approx 3~\text{mas}$ at $10~\text{\textmu m}$, would be possible. In our analysis we assumed a conservative IWA of $5.0~\text{mas}$ at $\lambda_\text{eff} = 10~\text{\textmu m}$. However, we also investigated the impact of shorter nulling baselines, i.e., worse IWA. For the OWA (outer working angle) we assumed $1~\lambda_\text{eff}/D$, where $D$ is the aperture diameter, which is $\sim1''$ at $\lambda_\text{eff} = 10~\text{\textmu m}$ for a $2.8~\text{m}$ aperture\footnote{To achieve the same collecting area as the \emph{JWST}, each of the four mirror spacecraft would need to have a diameter of $D \approx 2.8~\text{m}$.}. Both IWA and OWA are scaled with the corresponding effective wavelength $\lambda_\text{eff}$. \citet{cockell2009} further required the instrument to achieve a sensitivity comparable to the \emph{JWST}. In our analysis we focused on the F560W ($5.6~\text{\textmu m}$), F1000W ($10~\text{\textmu m}$), and F1500W ($15~\text{\textmu m}$) filters and their respective sensitivities from the mid-infrared instrument (MIRI) of the \emph{JWST}. We therefore used the MIRI filter curves and the faint source low background detection limits\footnote{They differ by less than $4\%$ from the high background values for our filters.} from Table~3 of \citet{glasse2015}, who performed an extensive study on the detector sensitivity taking into account zodiacal background at the L2 point, thermal instrument background, photon conversion efficiency, image broadening, scattering, and crosstalk. Since our three filters have effectively no overlap we assumed that all three bands can be observed simultaneously. However, comparing the instrument throughput of MIRI, which is approximately $35\%$ for the F1000W filter\footnote{\url{https://jwst.etc.stsci.edu/}; the F560W and the F1500W filters have lower throughputs.}, to those of a nulling interferometer \citep{dubovitsky2004}, which is approximately $10\%$ only, we scaled the necessary integration time for a $10\sigma$ detection by a factor of $3.5$, resulting in $35'000~\text{s}$.

The high spatial resolution in combination with the superb sensitivity is key for the direct detection of the thermal emission of close-in extrasolar planets. An exoplanet is considered as detectable in the simulations if its apparent angular separation exceeds the instrument's IWA and is smaller than the OWA and its total flux observed in a particular band exceeds $F_\text{lim}$ (cf. Table~\ref{baseline}). We do not consider null-depth in our simulations\footnote{\citet{cockell2009} require a null depth of $1\mathrm{e}{-5}$ stable over a duration of $5~\text{d}$.}, but discuss various noise effects in Section~\ref{observing_strategy}.


\subsection{Radial velocity detectability}
\label{radial_velocity}

In several cases we also analyzed the feasibility of additional radial velocity measurements. However, one needs to know the mass of the planet to estimate the semi-amplitude of the radial velocity signal
\begin{equation}
\label{eq05}
K_* = \sqrt{\frac{G}{1-e^2}}M_\text{p}|\sin(i)|(M_*+M_\text{p})^{-1/2}a^{-1/2},
\end{equation}
where $G$ is the gravitational constant, $e$ is the orbital eccentricity, $M_\text{p}$ is the mass of the planet, $i$ is the orbital inclination, $M_*$ is the host star mass, and $a$ is the semi-major axis of the orbit of the exoplanet around its host star. Having its radius already (simulated from the \emph{Kepler} planet occurrence statistics) and assuming a planet density of $\rho_\text{p} = 5'000~\text{kg}~\text{m}^{-3}$ for all planets with radii below $2~R_\text{Earth}$, we approximated the mass of the planet. The density is motivated by the mean value of the densities of the four terrestrial planets of our own solar system, which is $5'029~\text{kg}~\text{m}^{-3}$. However, the assumption that all planets with radii below $2~R_\text{Earth}$ are rocky is a rough simplification. The transition from rocky planets to planets with a significant gas envelope lies probably between $1.2$ and $1.8~R_\text{Earth}$ \citep{wolfgang2015}. Similarly, \citet{rogers2015} found that most planets with radii above $1.6~R_\text{Earth}$ are not rocky considering a \emph{Kepler} subsample of planets with radial velocity mass constraints. We thus keep in mind that our analysis might overestimate the radial velocity signal for a fraction of the exoplanets between $\sim1.5$ and $2~R_\text{Earth}$. As baseline detection threshold we adapted a value of $K_* \geq 10~\text{cm}~\text{s}^{-1}$ motivated by the aimed precision of the new Echelle spectrograph for rocky exoplanet and stable spectroscopic observations (ESPRESSO) currently installed at ESO's Very Large Telescope in Chile \citep{pepe2010}; we acknowledge that this precision is not necessarily be achieved for all stars in our sample. Especially, detecting exoplanets around A- and early F-type stars is challenging because of their sparse spectra and this might require additional efforts \citep[e.g., Fourier space correlation;][]{lagrange2009}.


\section{Results}

We present all of our results in terms of the expected number of detectable exoplanets ($\eta$). A real observation can be understood as a random experiment where a single value is drawn from a Poisson-like probability function with mean value $\eta$. All results are therefore affected by statistical Poisson errors $\Delta\eta$. Since we simulated $2'000$ exoplanetary systems around each host star these errors can be calculated for each expectation value of detectable exoplanets $\eta$ via
\begin{equation}
\label{eq06}
\Delta\eta = \frac{\sqrt{\eta \cdot 2000}}{2000} \approx 0.022 \cdot \sqrt{\eta}
\end{equation}
(error of a Poisson distribution).


\begin{figure}[h!]
\centering
\includegraphics[width=\hsize]{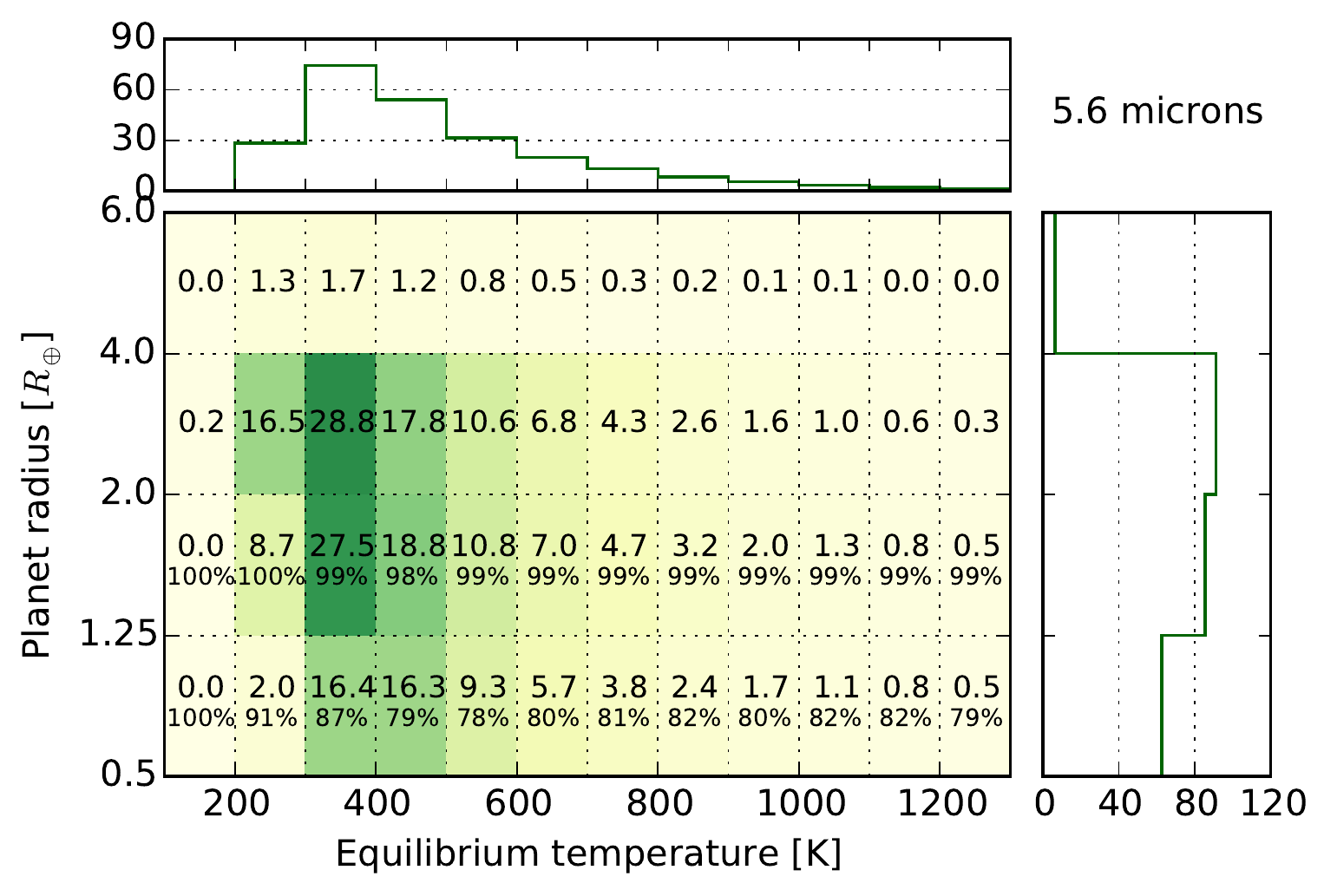}
\includegraphics[width=\hsize]{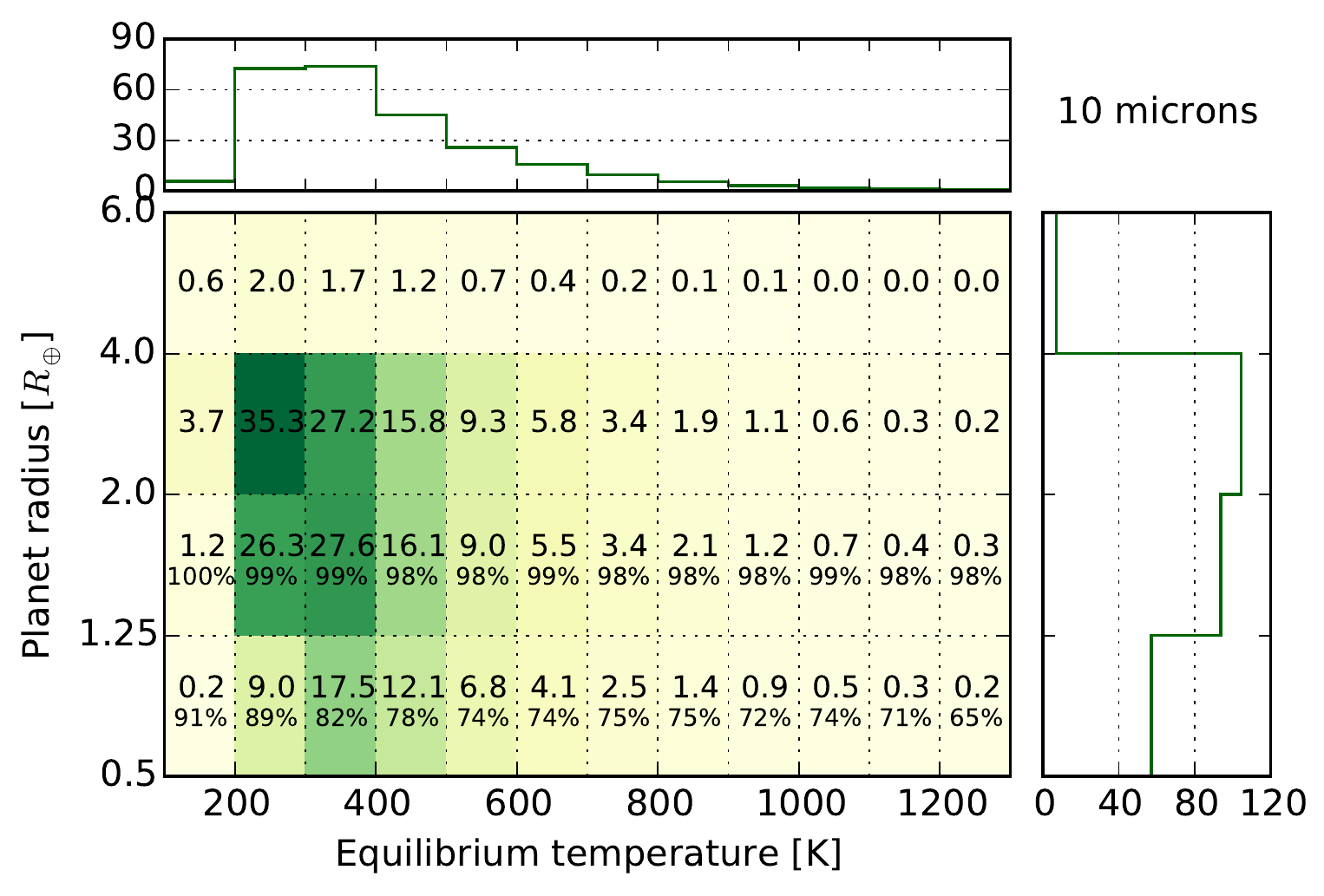}
\includegraphics[width=\hsize]{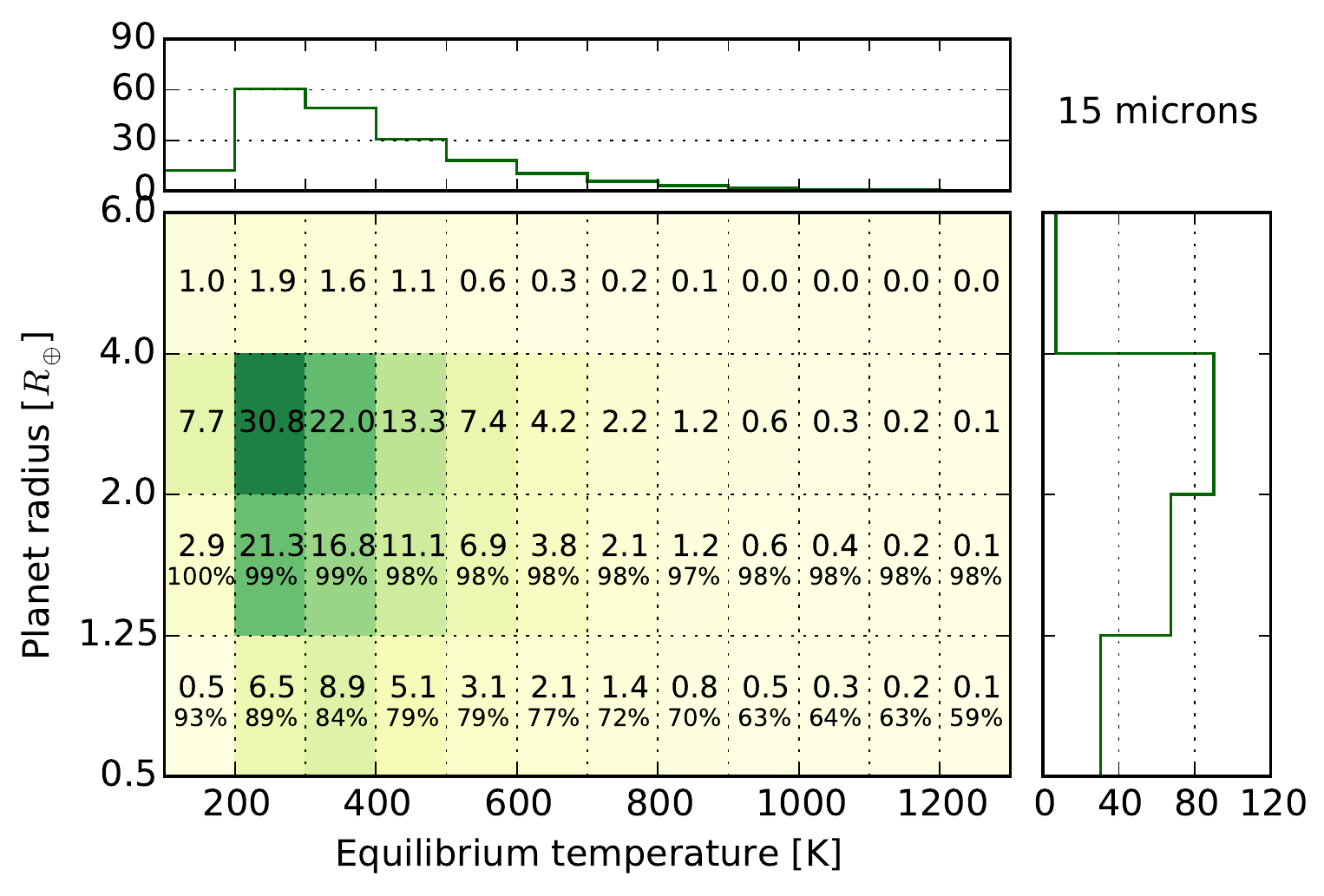}
\caption{Expected number of detectable exoplanets binned in the radius-equilibrium temperature plane for the $5.6~\text{\textmu m}$ filter (F560W, top),$10~\text{\textmu m}$ filter (F1000W, middle), and $15~\text{\textmu m}$ filter (F1500W, bottom). The histograms in each subfigure show the projected equilibrium temperature distribution (top) and the projected radius distribution (right) of the detectable exoplanets. The numbers in percent state the percentage of planets, which is also detectable with radial velocity observations assuming a detection threshold of $K_* \geq 10~\text{cm}~\text{s}^{-1}$ and a planet density of $\rho_\text{p} = 5'000~\text{kg}~\text{m}^{-3}$. Color shade and figure axes are scaled equally in all three subfigures for better comparison. A negligible percentage of detectable exoplanets ($\lessapprox 1\%$) have equilibrium temperatures below $100~\text{K}$ or above $1'300~\text{K}$ and thus lie outside the depicted bins.}
\label{fig01}
\end{figure}


\subsection{Single filter observations}


\subsubsection{Number of detectable exoplanets -- Baseline scenario}

The expected numbers of detectable exoplanets for each individual filter are given in Figure~\ref{fig01}. We restrict the presented parameter space to planets with radii up to $6~R_\text{Earth}$ to stick to the planet radius bins determined by \citet{fressin2013}\footnote{The occurrence rate of planets with radii between $6$ and $22~R_\text{Earth}$ is comparatively low.}. Integrating over the whole depicted radius-equilibrium temperature plane results in $\sim243$ expected planets at $5.6~\text{\textmu m}$, $\sim261$ expected planets at $10~\text{\textmu m,}$ and $\sim194$ expected planets at $15~\text{\textmu m}$. The majority of these planets are smaller than Neptune ($R_\text{p} \leq 4~R_\text{Earth}$) and have equilibrium temperatures between $200$ and $700~\text{K}$.

The highest number of exoplanets can be detected with the $10~\text{\textmu m}$ filter although it is inferior to the $5.6~\text{\textmu m}$ filter in terms of spatial resolution (i.e., IWA and scales with $\lambda_\text{eff}$) and sensitivity. However, the simulated planet population consists of a significant number of planets with equilibrium temperatures around $300~\text{K}$. Their thermal emission peaks at roughly $10~\text{\textmu m}$, but their flux decreases sharply toward shorter wavelengths so that they remain undetected in the $5.6~\text{\textmu m}$ filter. Going from shorter to longer wavelengths shows that the average equilibrium temperatures of the detectable exoplanets shift slightly toward cooler planets because the planet equilibrium temperature is proportional to the inverse square root of the physical separation between planet and host star; hotter planets are located closer to their host stars and cannot be resolved at longer wavelengths.

The percentage of exoplanets that can also be detected with radial velocity measurements (as defined in Section~\ref{radial_velocity}) is very high for planets between $1.25$ and $2~R_\text{Earth}$, typically close to $100\%$. For these planets the true mass and orbit could be reconstructed by combining data from both observing techniques, including a constraint on the radii of the planets from their measured luminosity (provided that their effective temperature can be reliably estimated). Because of their expected lower mass, planets with radii between $0.5$ and $1.25~R_\text{Earth}$ are less frequently detectable with radial velocity. However, the percentage detected of such planets is still mostly above $70\%$.


\subsubsection{Instrument performance}

\begin{figure}[h!]
\centering
\includegraphics[width=\hsize]{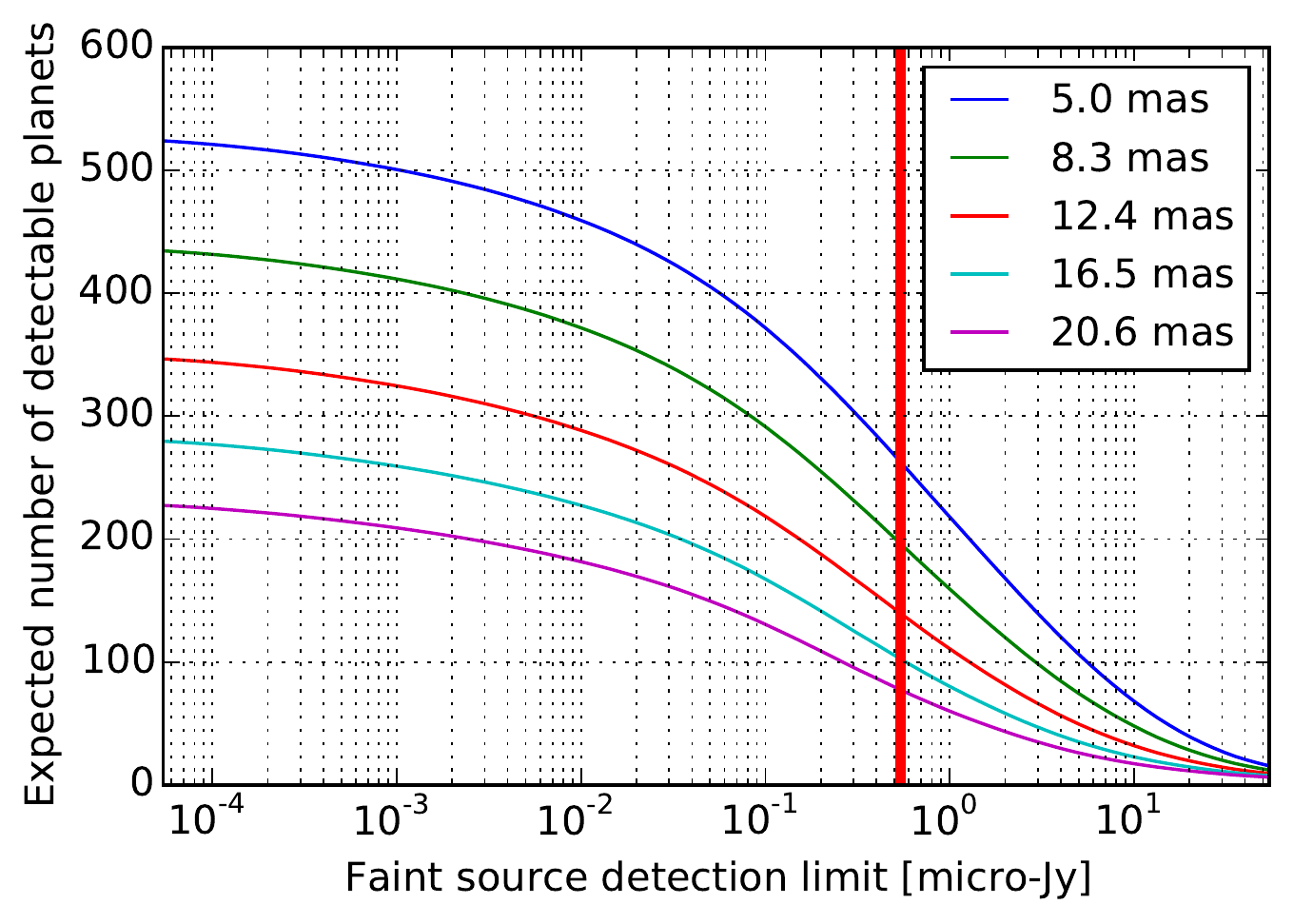}
\includegraphics[width=\hsize]{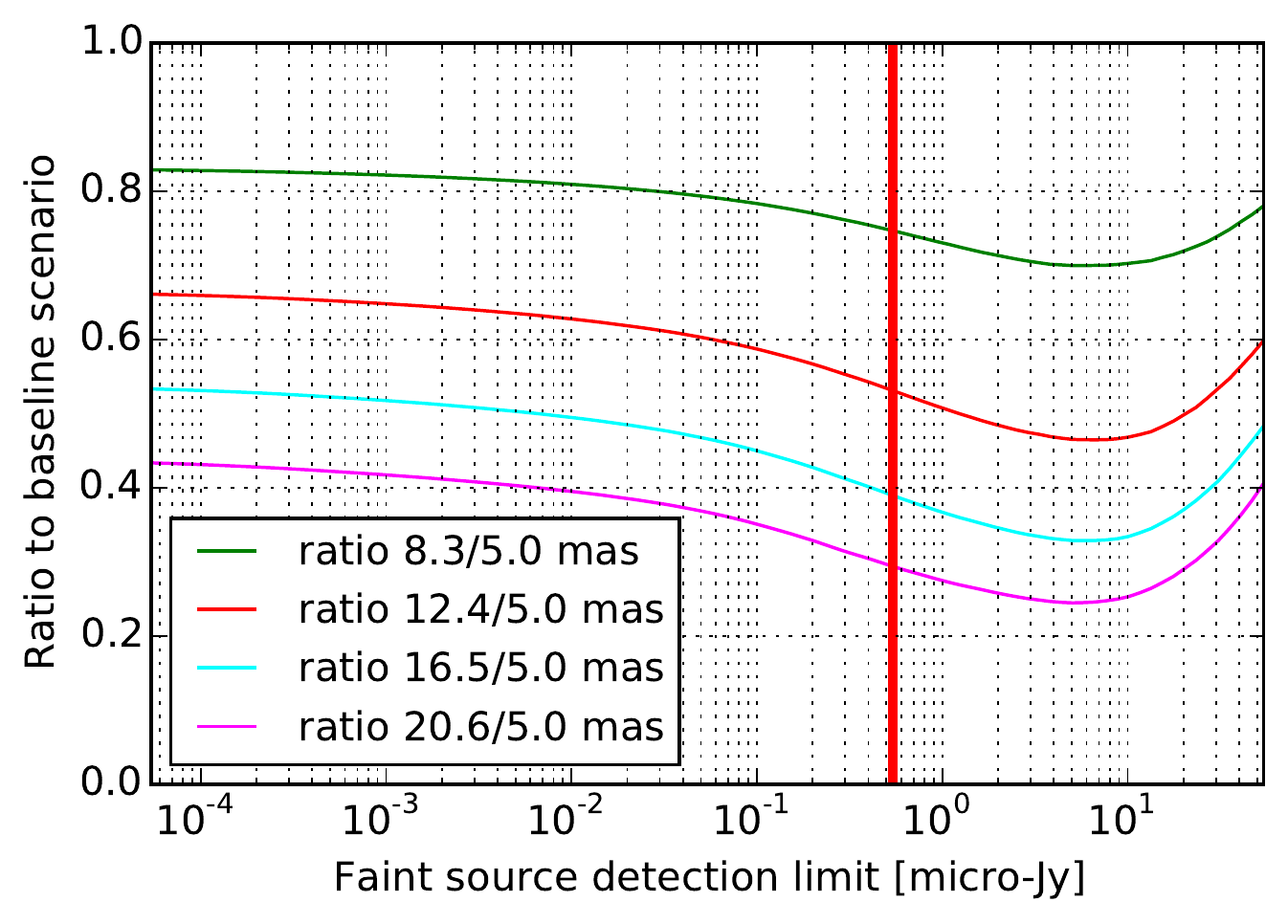}
\caption{Top: Expected number of detectable exoplanets (with radii $< 6~R_\text{Earth}$) as a function of the faint source detection limit $F_\text{lim}$ for the $10~\text{\textmu m}$ filter. The five curves represent various nulling baselines, i.e., various IWAs ranging from $5.0~\text{mas}$ (baseline value) to $20.6~\text{mas}$. The red vertical line indicates our baseline detection limit of $F_\text{lim,~F1000W} = 0.54~\text{\textmu Jy}$. Bottom: Ratios of the other curves to the $5.0~\text{mas}$ curve as a function of the faint source detection limit $F_\text{lim}$.}
\label{fig02}
\end{figure}

We investigated how sensitive the scientific output, i.e., the expected number of detectable exoplanets, depends on the assumed instrument performance focusing on the $10~\text{\textmu m}$ filter because it promises the highest planet yield in our baseline scenario. Figure~\ref{fig02} (top panel) shows the expected number of detectable exoplanets as a function of the assumed faint source detection limit $F_\text{lim}$ for five different nulling baselines resulting in IWAs ranging from $5.0$ to $20.6~\text{mas}$. The slopes of the individual curves are a measure of how stable the expected exoplanet yield is with respect to the sensitivity of the instrument. For all presented IWAs the curves have their steepest slope at roughly our baseline scenario of $F_\text{lim,~F1000W } = 0.54~\text{\textmu Jy}$ implying that the instrument is operating in a regime where its scientific yield is particularly sensitive to variations in the achievable detection limit (see also Section~\ref{observing_strategy}).


\begin{figure*}[h!]
\centering
\includegraphics[width=9cm]{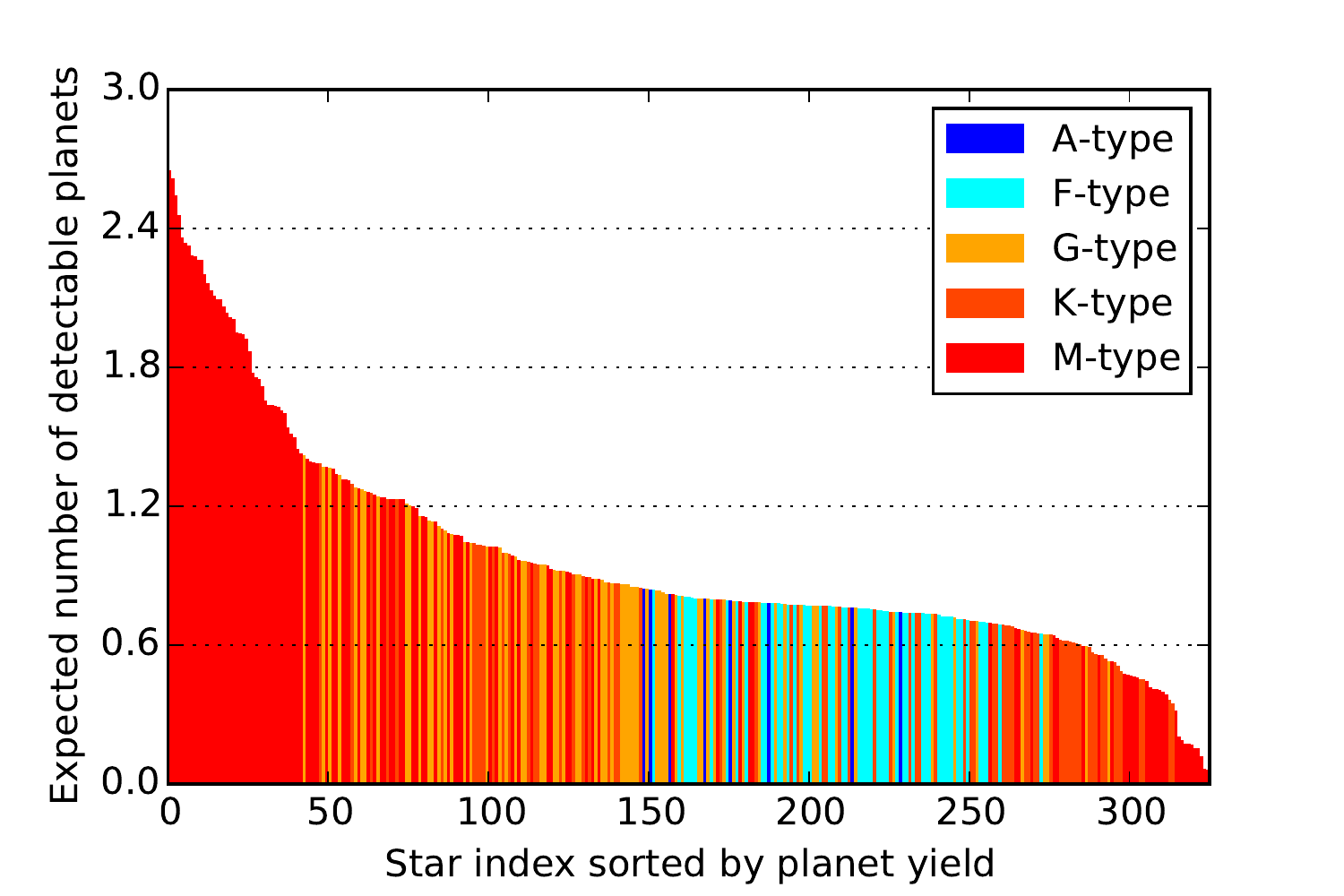}
\includegraphics[width=9cm]{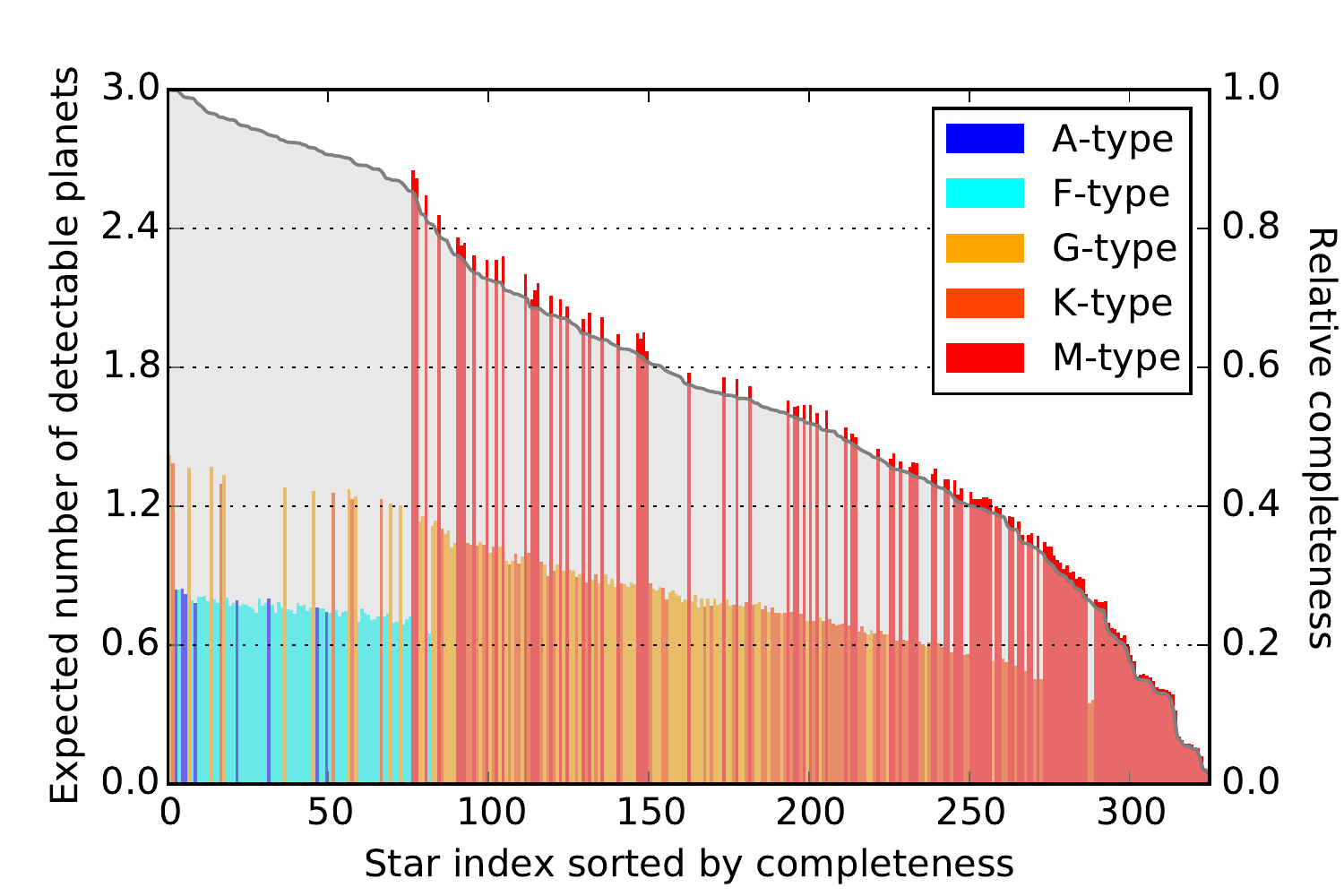}
\caption{Left: Expected number of detectable exoplanets (with radii $< 6~R_\text{Earth}$) for each individual host star from our star catalog. Each of the $326$ host stars is represented by one bin of the histogram (see color code for the different spectral types in the figure legend). This plot contains all exoplanets that can be detected in at least one of the three filters used in this work ($5.6~\text{\textmu m}$, $10~\text{\textmu m}$, $15~\text{\textmu m}$). Right: Expected number of detectable exoplanets for each individual host star from our star catalog (left axis, as in the left panel), but sorted by completeness (right axis) overplotted as the transparent gray curve on the histogram.}
\label{fig03}
\end{figure*}


\begin{figure}[h!]
\centering
\includegraphics[width=9.0cm]{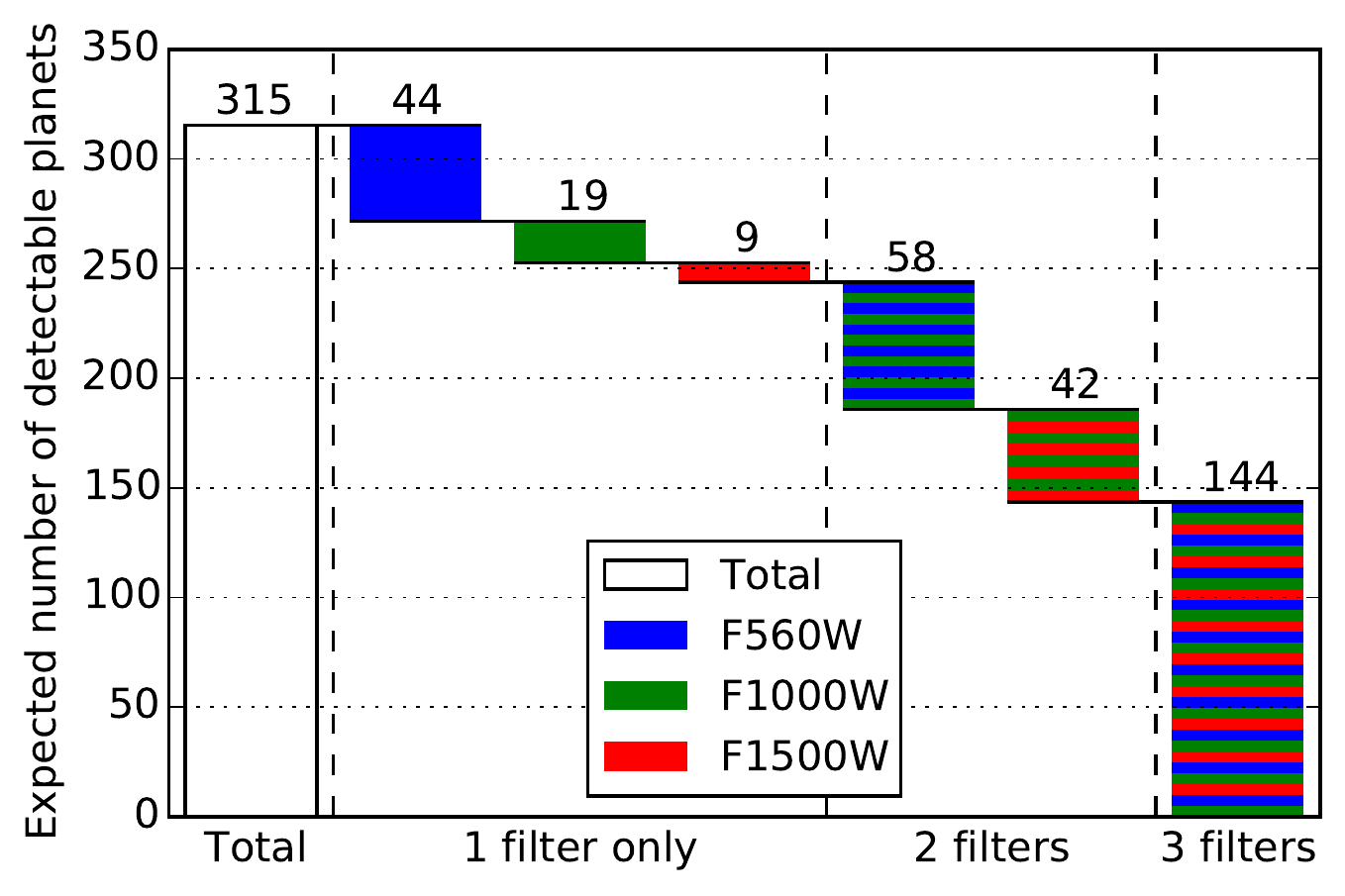}
\caption{Expected number of exoplanets (with radii $< 6~R_\text{Earth}$) that are detectable in only one, two, or all three filters (see color code for the different filters in the figure legend). There are no planets expected to be detectable simultaneously at $5.6$ and $15~\text{\textmu m}$ alone. In total $\sim315$ exoplanets can be detected in at least one of our simulated filters. Any number of discrepancies are due to rounding errors.}
\label{fig04}
\end{figure}


The ratios of the expected exoplanet yield of the four smaller nulling baselines, i.e., worse IWAs, relative to our baseline value of $5.0~\text{mas}$ are plotted in the bottom panel of Figure~\ref{fig02}. This figure shows that these lines are roughly constant, in particular the green line that depicts the ratio between $8.3$ and $5.0~\text{mas}$, for detection limits that are more sensitive than our baseline value. In this regime, for an IWA of $8.3~\text{mas}$, the planet yield is $\sim20\%$ lower; for an IWA of $20.6~\text{mas}$ only $\sim40\%$ of the planets are detected compared to our baseline scenario.

Toward the right side of the presented parameter space all four curves in the bottom panel of Figure~\ref{fig02} show a small dip. This implies that the expected exoplanet yield is most sensitive to variations in the IWA toward faint source detection limits roughly one order of magnitude worse than our baseline assumption.


\subsection{Multi-filter observations}


\subsubsection{Spectroscopic characterization}

For a more comprehensive characterization of their properties (e.g., atmospheric composition) the exoplanets should be analyzed spectroscopically and hence they must be detectable over a broad wavelength range in a reasonable amount of time. To estimate the percentage of exoplanets for which a spectroscopic characterization would be possible Figure~\ref{fig04} shows the expected number of exoplanets that can be detected in only one, two, or in all three filters. Out of the expected $\sim315$ individual exoplanets, which can be detected in at least one filter, $\sim144$ exoplanets can be detected in all three filters.

Therefore, as a significant percentage of exoplanets are detected between $5.6$ and $15~\text{\textmu m}$ within a reasonable amount of observing time, there would be sufficient time left for detailed follow-up observations of the most interesting targets. The MIR wavelength regime contains numerous absorption bands of key molecules for atmospheric characterization \citep[e.g., $\text{O}_3$, at $9.6~\text{\textmu m}$; $\text{CO}_2$, at $15~\text{\textmu m}$; $\text{H}_2\text{O}$, at $7$ and $20.5~\text{\textmu m}$; and $\text{CH}_4$, at $3.3$ and $8~\text{\textmu m}$; cf.][]{desmarais2002} and bands from additional biosignatures \citep[such as $\text{CH}_3\text{Cl}$, $\text{DMS}$, $\text{N}_2$, and $\text{NH}_3$; cf.][]{seager2013}. A detailed analysis concerning the required S/N, spectral resolution, and wavelength range to identify certain molecules and robustly characterize atmospheric properties as a function of planet parameters is, however, beyond the scope of the present paper and left for future investigations.


\subsubsection{Planet yield per star}

\begin{table}[h!]
\caption[]{\label{yield} Name, distance $d$, spectral type, completeness $C$, and expected number of detectable exoplanets (planet yield) for the best targets.}
\centering
\begin{tabular}{lllll}
\hline\hline
\noalign{\smallskip}
Name           & $d$ [pc] & Spec. type  & $C$  & Planet yield \\
\noalign{\smallskip}
\hline
\noalign{\smallskip}
Gl 887         & 3.28     & M2V         & 0.85 & 2.7 \\
Gl 15 A        & 3.59     & M1.0V       & 0.85 & 2.6 \\
Gl 411         & 2.55     & M1.5V       & 0.82 & 2.5 \\
Gl 1           & 4.34     & M2V         & 0.79 & 2.5 \\
Gl 725 A       & 3.57     & M3.0V       & 0.76 & 2.4 \\
Gl 338 A       & 6.14     & M0.0V       & 0.76 & 2.3 \\
Gl 832         & 4.95     & M2/3V       & 0.76 & 2.3 \\
Gl 784         & 6.20     & M0V         & 0.74 & 2.3 \\
Gl 526         & 5.39     & M3V         & 0.72 & 2.3 \\
Gl 628         & 4.29     & M3V         & 0.73 & 2.3 \\
\noalign{\smallskip}
\hline
\noalign{\smallskip}
$\alpha$ Cen A & 1.32     & G2V         & 1.00 & 1.4 \\
$\alpha$ Cen B & 1.32     & K1V         & 1.00 & 1.4 \\
Sirius A       & 2.64     & A1V         & 1.00 & 0.8 \\
Procyon A      & 3.51     & F5IV-V      & 1.00 & 0.8 \\
Altair         & 5.13     & A7Vn        & 0.99 & 0.9 \\
Fomalhaut      & 7.70     & A4V         & 0.99 & 0.8 \\
$\delta$ Pav   & 6.11     & G8IV        & 0.99 & 1.4 \\
$\eta$ Cas A   & 5.95     & F9V         & 0.99 & 0.8 \\
Vega           & 7.68     & A0Va        & 0.99 & 0.8 \\
$\zeta$ Tuc    & 8.59     & F9.5V       & 0.98 & 0.8 \\
\noalign{\smallskip}
\hline
\end{tabular}
\tablefoot{The top half of the table shows the best 10 targets according to the left panel of Figure~\ref{fig03} (highest planet yield), and the bottom half shows the best 10 targets according to the right panel of Figure~\ref{fig03} (highest completeness). Spectral types taken from the SIMBAD Astronomical Database (\url{http://simbad.u-strasbg.fr/simbad/}).
}
\end{table}

From the point of view of the observer, it is not only interesting to know how many exoplanets one can expect to detect, but also around which targets one could find these exoplanets. For this purpose we summarize the expected number of detectable exoplanets for each individual host star from our star catalog in Figure~\ref{fig03} (left panel). We sort the stars by planet yield to show that most of the stars have expectation values of $\sim1$ detectable exoplanet. For roughly $30\%$ of the host stars from our star catalog our simulations predict an expected planet yield of $\geq 1$, ultimately ranging up to $\sim2.7$.

The top half of Table~\ref{yield} reveals that the 10 stars with the highest expected exoplanet yield are all nearby M-type stars. However, the exoplanet yield alone does not quantify the detection completeness, i.e., what percentage of the generated planets in each individual exoplanetary system is detectable. To decouple the underlying planet population from our analysis, we calculated the completeness for each host star as the fraction of exoplanets that can be directly detected relative to all exoplanets that are simulated around this star. Figure~\ref{fig03} (right panel) shows the expected number of detectable exoplanets for each target (like in the left panel), but this time the stars are sorted by their completeness rather than their expected exoplanet yield. The bottom half of Table~\ref{yield} reveals that the completeness is close to $100\%$ for nearby solar-type and intermediate-mass stars. This trend of higher completeness toward earlier spectral types can be explained with the comparatively higher temperatures of their planets.


\subsection{Distribution of spectral types}

\begin{table}[h!]
\caption{\label{types}Total expected exoplanet yield and expected exoplanet yield per star and average completeness $C$ sorted by host star spectral type.}
\centering
\begin{tabular}{llll}
\hline\hline
\noalign{\smallskip}
\multirow{2}{*}{Spec. type} & \multicolumn{2}{l}{Planet yield} & \multirow{2}{*}{$C$} \\
                            & total & per star                 & \\
\noalign{\smallskip}
\hline
\noalign{\smallskip}
A                         & 6.4   & 0.80                     & 0.96 \\
F                         & 40.8  & 0.76                     & 0.92 \\
G                         & 66.4  & 0.92                     & 0.67 \\
K                         & 55.0  & 0.78                     & 0.56 \\
M                         & 146.8 & 1.21                     & 0.39 \\
\noalign{\smallskip}
\hline
\end{tabular}
\end{table}

In case one aims at comparing the properties of exoplanets as a function of their host star spectral type, for example, from a planet formation perspective, Table~\ref{types} shows that planets (with radii $< 6~R_\text{Earth}$) around all types of host stars are detected in our baseline scenario. Roughly one-half of the detectable planets orbit M stars and the other half orbit FGK stars; A stars, of which we have only eight in our sample, are negligible.


\section{Discussion}

As motivated in the Introduction the goal of this paper is to quantify the expected exoplanet yield of a space-based nulling interferometer to provide a framework for deriving robust science requirements for a possible future mission. Therefore, we discuss in the following how the expected exoplanet yield depends on our astrophysical assumptions such as planetary albedos, orbital eccentricity, and (exo-)zodiacal light. We further investigate the planet occurrence rate around binary stars, assess the impact of the statistical errors in the underlying planet population, review our observing strategy in the context of stellar leakage, and discuss the stellar properties of our M dwarf host stars. We end our discussion section by specifying the expected exo-Earth yield of our nulling interferometer in comparison to that of a large space-based optical/NIR telescope such as proposed in the NASA \emph{HabEx} and \emph{LUVOIR} missions.


\subsection{Choice of albedos}
\label{albedos}

The expected number of detectable exoplanets is strongly dependent on the choice of the planetary albedos, particularly on the choice of the Bond albedos. \citet{crossfield2013} and \citet{quanz2015} assume a uniform distribution of the Bond (and the geometric) albedos between $0$ and $0.4$. However, considering the Bond albedos of the solar system planets, we distributed these albedos uniformly between $0$ and $0.8$ for our simulated planets. This increase in the Bond albedo range reduces the scientific yield since higher Bond albedos translate into lower equilibrium temperatures and therefore less thermal emission.

A reasonable choice of the geometric albedos is more difficult since they are wavelength dependent, but hardly any information is available for the wavelength range we are interested in. We decided to distribute the geometric albedos uniformly between $0$ and $0.1$ because many gases that dominate the atmospheres of our solar system planets are strongly absorbed in the MIR. However, we find that in our simulations the contribution of the reflected host star light is negligible for the detection of the vast majority of exoplanets. Switching off the reflected host star light contribution reduces the expected number of detectable exoplanets in the $5.6~\text{\textmu m}$ filter, where one would expect the strongest effect, by no more than $\sim1\%$. On the other hand, even if we distributed the geometric albedos uniformly between $0$ and $1$ instead of $0$ and $0.1$, the expected exoplanet yield would increase by no more than $\sim1\%$. More specifically, the expected number of detectable exoplanets would increase by $\sim4.4$ planets with equilibrium temperatures $T_\text{eq,~p} \leq 300~\text{K}$. Therefore, the choice of the geometric albedos has only a minor impact on our results.


\subsection{Choice of eccentricity}
\label{eccentricity}

For our baseline scenario we assumed circular orbits (eccentricity $e = 0$; see Table~\ref{baseline}), however our Monte Carlo simulations can also handle eccentric orbits. On the one hand, planets on eccentric orbits spend more time around their apoapsis where the separation between exoplanet and host star is largest. Therefore, one would expect more detections because (statistically) more exoplanets can be spatially resolved from their host stars. On the other hand, an exoplanet that is located further away from its host star will be cooler and emit less thermal blackbody radiation and less reflected light. Therefore one would expect less detections because (statistically) less exoplanets pass the detection threshold of the instrument. Here we assumed that the timescales of the thermodynamic processes in the atmospheres of the exoplanets are short compared to their orbital periods so that the equilibrium temperatures of the exoplanets are always determined by the instantaneous physical separations between exoplanets and host stars $r_\text{p}$. Comparing the expected number of detectable exoplanets for circular orbits ($e = 0$) and highly eccentric orbits ($e$ distributed uniformly in $[0,~1)$) it can be concluded that the second effect has a larger impact since in our simulations only $\sim293$ are expected in the case of highly eccentric orbits compared to $\sim315$ in the non-eccentric case. However, as the difference is small (roughly $-7\%$) we consider our results as robust with respect to variations in the orbital eccentricity. If we assume a delay in the cooling of the exoplanets whilst moving from smaller to larger physical separations from their host stars, the exoplanets still had a higher temperature and emitted more thermal radiation when they reach their apoapsis. This would lead to an increase in the expected exoplanet yield, but only of the same order of magnitude as the decrease assuming instantaneous cooling and heating.


\subsection{(Exo-)zodiacal Light}

Planetary systems can be filled with dust particles, which do not only reflect stellar light but also emit thermal radiation. In our solar system this emission is called zodiacal light; around other stars it is referred to as exozodiacal light. This emission, from within our own solar system, but also from within the other stellar systems, gives rise to a significant radiation background in the thermal infrared. In fact, there are several projects ongoing to characterize the exozodiacal light in nearby stellar systems in preparation for future planet searches \citep[e.g.,][]{danchi2016, defrere2016}.

In our simulations the impact of the zodiacal light on our faint source detection limits is fully taken into account because the scattered and emissive components of the interplanetary dust are included in the MIRI sensitivity model of \citet{glasse2015}. These authors predicted that zodiacal light is the dominant background source shortward of $\sim17~\text{\textmu m}$ and the reflected and thermal emission of the instrument becomes dominant longward of $\sim17~\text{\textmu m}$ only. The zodiacal light contribution depends on the orbit of the observatory, but we assume that our MIR interferometer is operated at the Earth-Sun L2 point similar to the \emph{JWST}.

The contamination from the exozodiacal light is not taken into account in our simulations. If these emissions were uniform and without significant structure, for example, from a uniform face-on dust disk, then they could be canceled out by the nulling interferometer. According to \citet{cockell2009}, by chopping the outputs of two Bracewell interferometers via shifting their phase by $\pm\pi/2$, centro-symmetric sources can be suppressed significantly. This procedure not only eliminates exozodiacal light but also stellar leakage from the host star. Yet, even if the emission itself can be canceled out, there will still be an increase in background noise and hence a loss in sensitivity. For an exozodiacal dust cloud, which is less than five times as dense as our solar system zodiacal cloud, \citet{dubovitsky2004} find that its photon noise is negligible compared to that originating from stellar leakage (see Section~\ref{observing_strategy}) and zodiacal light. However, at higher dust densities, a significant portion of the photon noise can originate from the exozodiacal dust cloud so that our simulations overestimate the number of detectable exoplanets in this case.

While including estimations for exozodiacal light in our simulations is certainly one of the next steps, we point out that, as discussed in Section~\ref{observing_strategy}, at the moment we assume a constant integration time of $35'000~\text{s}$ per star in our baseline scenario and we could easily increase (and optimize) the observing time for individual targets.


\subsection{Planet occurrence around binaries}

Our simulations predict the number of detectable exoplanets within a distance of $20~\text{pc}$ only excluding close binaries. However, many processes, such as disk truncation \citep{jang-condell2015}, enhanced accretion \citep{jensen2007}, and enhanced photoevaporation \citep{alexander2012}, can influence the formation and evolution of planets in binary star systems. \cite{kraus2016} have shown that the occurrence of planets around close-in binaries with separations below $a_\text{cut} = 47_{-23}^{+59}~\text{au}$ differs from that around single stars or wider separated binaries by a suppression factor of $S_\text{cut} = 0.34_{-0.15}^{+0.14}$. They have used high-resolution imaging follow-up observations of a subsample of $382$ Kepler Objects of Interest (KOIs) to distinguish single star hosts from multiple star hosts. Since the \emph{Kepler} target stars were chosen almost blindly with respect to stellar multiplicity \citep{kraus2016} their findings translate into a higher planet occurrence for a stellar sample that systematically excludes close-in binaries.

Assuming a Gaussian distribution of binary companions in $\log(P)$ space according to \cite{raghavan2010}, we find that $f_{< 47~\text{au}} = 49_{-7}^{+9}\%$ of all binary stars have separations below $a_\text{cut}$. Moreover, we find a scaling factor of $r_\text{single}/r = 1.14_{-0.07}^{+0.07}$ for the planet occurrence around single stars or wider separated binaries via
\begin{equation}
S_\text{cut} \cdot f_\text{bin} \cdot f_{< 47~\text{au}} \cdot r_\text{single} + (1-f_\text{bin} \cdot f_{< 47~\text{au}}) \cdot r_\text{single} = r,
\end{equation}
where $f_\text{bin}$ is the percentage of binary stars among \emph{Kepler} targets and $r$ is the occurrence rate of planets predicted by the \emph{Kepler} data. We took $f_\text{bin}$ from Table~16 of \cite{raghavan2010} who investigated a sample of $454$ solar-type stars for stellar companions ($f_\text{bin} = N_\text{bin}/(N_\text{single}+N_\text{bin})$). Their findings are valid for stellar masses between roughly $0.5~M_\odot \leq M_* \leq 1.5~M_\odot$. This means that a similar scaling factor could be calculated for the low-mass stars from our star catalog on the basis of a different distribution of binary companions \citep[e.g.,][]{janson2012}.

However, we decided not to scale our planet occurrence rates with this factor of $r_\text{single}/r = 1.14_{-0.07}^{+0.07}$ and regarded the total expected exoplanet yield as conservative with respect to stellar multiplicity. Nevertheless, we need to point out that some of the best targets (e.g., $\alpha$ Cen A/B, Sirius A, Procyon A) are in  binary systems with separations below $a_\text{cut}$ to which the suppressed planet occurrence rate would apply.


\subsection{Underlying planet population}

The underlying planet population has a high impact on the properties of the detectable exoplanets. The peak observed at radii of $1.25~R_\text{Earth} \leq R_\text{p} \leq 4~R_\text{Earth}$ and equilibrium temperatures of $200~\text{K} \leq T_\text{eq,~p} \leq 500~\text{K}$ is a direct fingerprint of the planet radius and orbital period distribution used to generate the exoplanetary systems. The spatial resolution and sensitivity of our space-based MIR interferometer allow for the detection of a broad range of planet types so that the underlying planet population itself is, to some extent, a limiting factor as also shown by the completeness analysis above. Another important point is our patchwork-like approach to merge planet occurrence statistics from different authors. Each statistic rests on different assumptions and models to estimate the completeness of the \emph{Kepler} pipeline and infer the true population of exoplanets from the highly biased sample of transiting exoplanets. We implicitly assumed a spectral type dependence of the planet occurrence by applying the statistics from \citet{dressing2015} to M-type stars only and those from \citet{burke2015} to G- and K-type stars only. This choice is reflected in our results (Table~\ref{yield} and Table~\ref{types}) as the targets with the highest planet yield are M-type stars.


\begin{figure}[h!]
\centering
\includegraphics[width=\hsize]{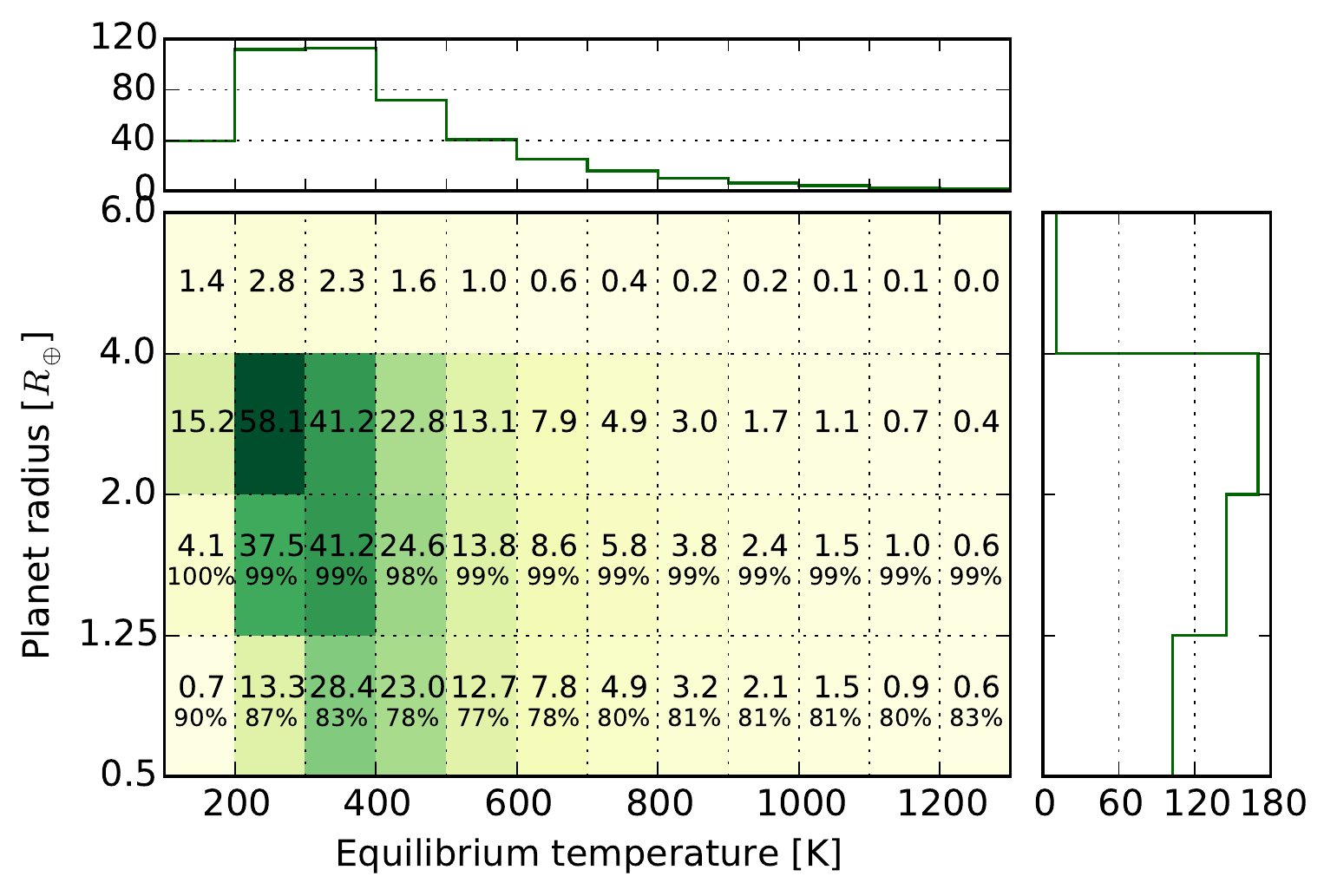}
\includegraphics[width=\hsize]{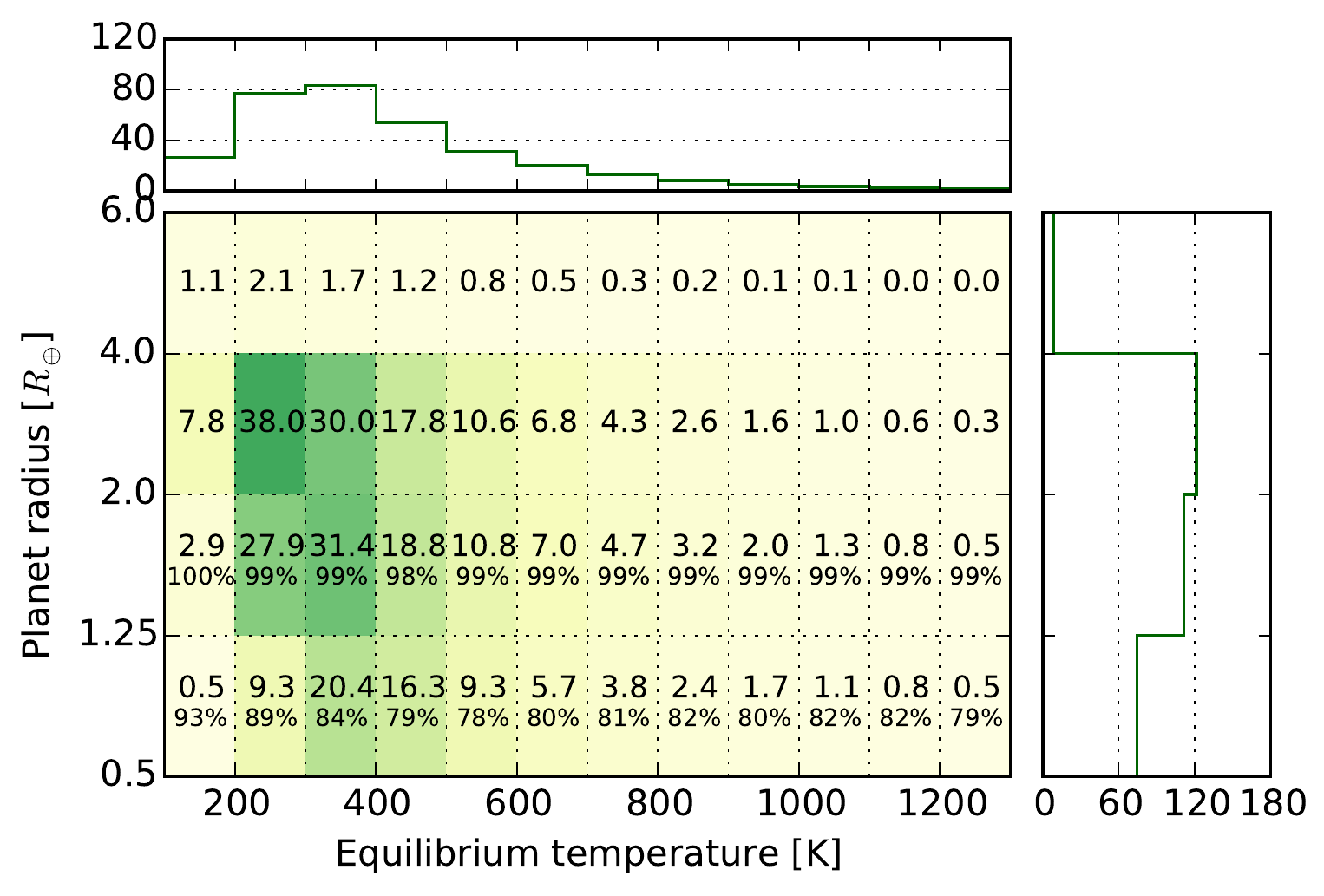}
\includegraphics[width=\hsize]{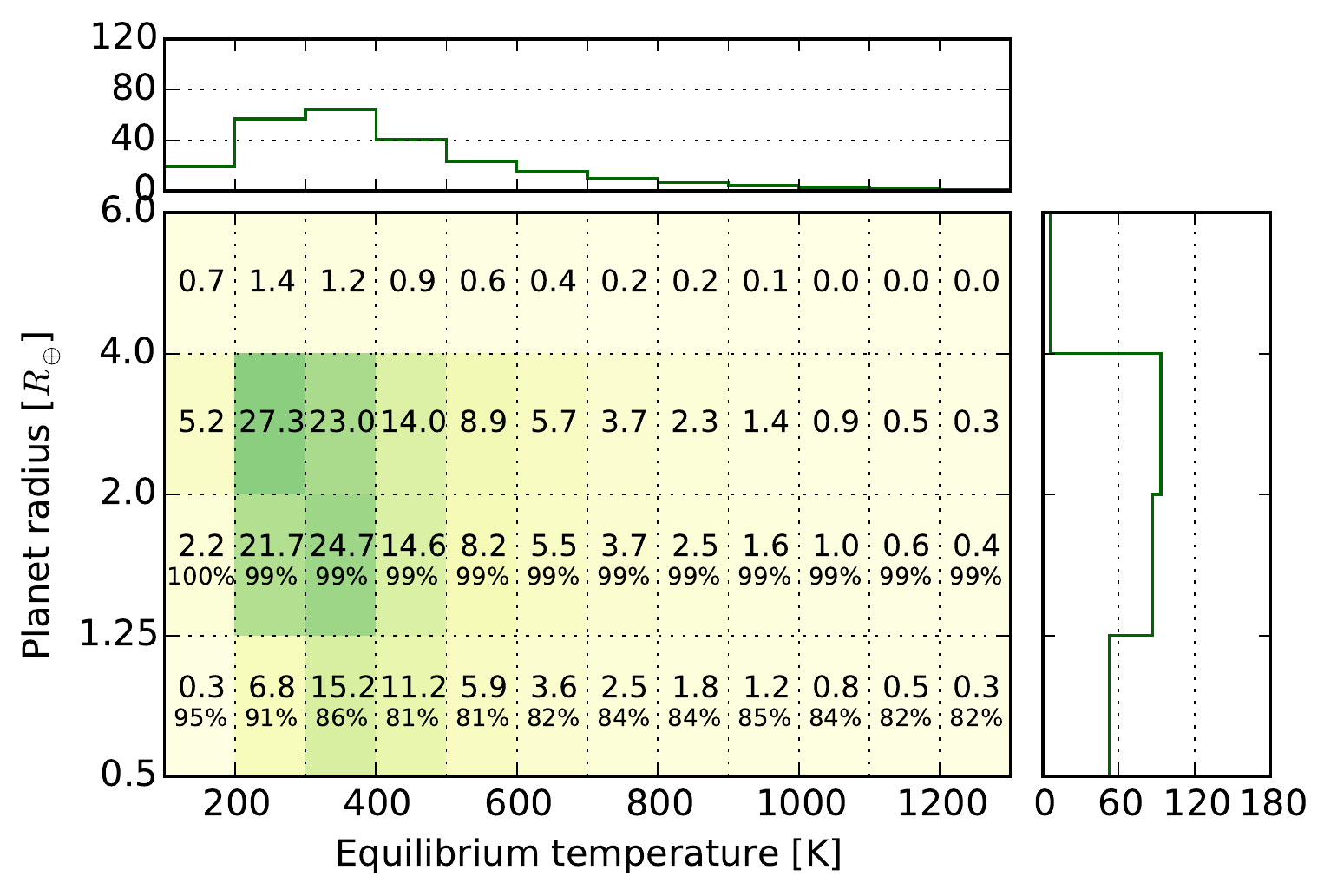}
\caption{Expected number of detectable exoplanets combined for all three simulated filters for our best-case scenario (top; $r+\Delta r$), for our baseline scenario (middle), and for our worst-case scenario (bottom; $r-\Delta r$), binned in the radius-equilibrium temperature plane (like in Figure~\ref{fig01})}
\label{fig05}
\end{figure}


Moreover, the underlying planet population is affected by statistical errors $\Delta r$. \citet{dressing2015} and \citet{fressin2013} presented these errors as upper ($r+\Delta r$) and lower ($r-\Delta r$) bounds for the planet occurrence rate $r$ within each individual planet radius-orbital period bin. \citet{burke2015} present an optimistic and a pessimistic efficiency scenario for their two-power-law model. We performed error analyses by comparing the expected exoplanet yield $r$ of our baseline scenario with that of best-case and worst-case scenarios with planet occurrence rates  $r+\Delta r$ and $r-\Delta r$, respectively (Figure~\ref{fig05}). We find an expected exoplanet yield of $\sim428$ in the best-case and $\sim238$ in the worst-case scenario revealing that the Poisson errors $315.4_{-0.4}^{+0.4}$ from our Monte Carlo simulation (cf. Equation~\ref{eq06}) are negligible compared to the statistical errors in the underlying planet population. Hence, we finally state the expected number of detectable exoplanets for our simulated mission as $315_{-77}^{+113}$.


\subsection{Observing strategy and integration time per target}
\label{observing_strategy}

The mission schedule for a space-based interferometer would very likely consist of two phases: (1) a detection phase to identify as many exoplanets as possible and (2) a characterization phase to investigate the atmospheric properties of a promising subset of those planets in detail. In our simulations we are primarily concerned with the first phase, which we assume to be background-limited. This means that an exoplanet is considered to be detectable as soon as its observed flux exceeds the faint source detection limit of the instrument and its apparent angular separation from the host star exceeds the IWA and is smaller than the OWA of the instrument. In our calculations, we take into account explicitly neither the planet-host star flux contrast nor additional photon noise from exozodiacal light or stellar leakage. The detection limits from \citet{glasse2015} are given for a $10\sigma$ detection significance in $10'000~\text{s}$ observing time. Taking into account the roughly $3.5$ times worse throughput of a nulling interferometer \citep{dubovitsky2004} if compared to MIRI\footnote{\url{https://jwst.etc.stsci.edu/}} and overheads of $\sim40\%$\footnote{\citet{cockell2009} assume that $70\%$ of the mission lifetime is spent taking data what results in overheads of $0.3/0.7 \approx 40\%$ if compared to the pure observing time.}, this would translate into a total of $\sim189~\text{d}$ or $\sim0.52$ years if all three bands could be observed simultaneously.

However, according to \citet{dubovitsky2004}, photon noise from stellar leakage can be a serious issue for stars closer than $\sim15~\text{pc}$. Their Figure~8 shows that the integration times necessary to detect an Earth twin around some of the nearby stars can increase by a factor of up to $\sim10$ owing to stellar leakage. Based on this it appears more realistic that the detection phase of our simulated mission will last somewhat between two and three years, which is slightly more than the two years proposed by \citet{cockell2009}. While beyond the scope of this paper, this clearly shows that detailed investigations regarding an optimized observing strategy, including quantitative estimates for stellar leakage, are warranted.

\begin{figure}
\centering
\includegraphics[width=\hsize]{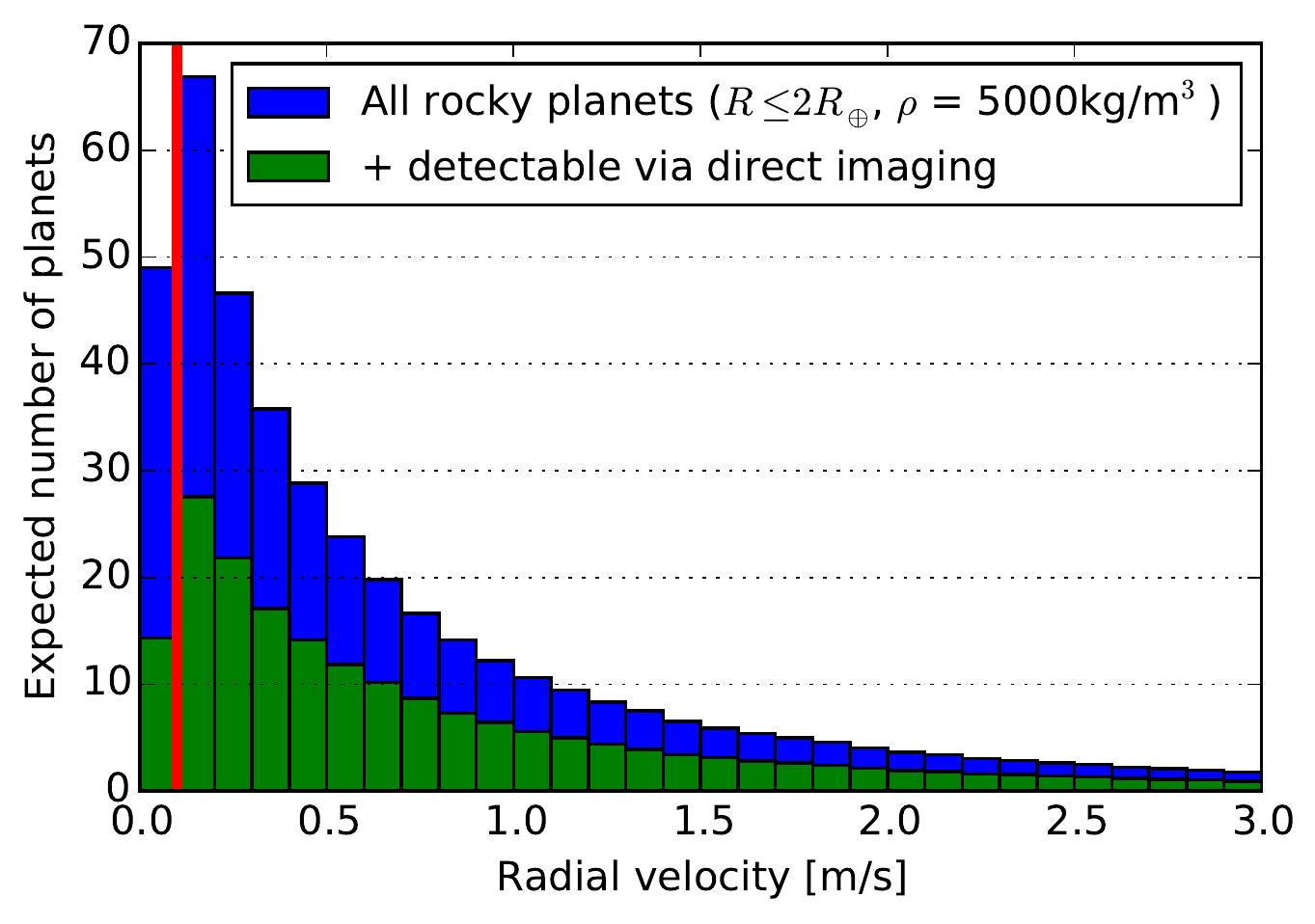}
\caption{Expected number of exoplanets binned by radial velocity semi-amplitude $K_*$ induced on their host star for all rocky exoplanets ($R_\text{p} \leq 2~R_\text{Earth}$) assuming a planet density of $\rho_\text{p} = 5'000~\text{kg}~\text{m}^{-3}$ (blue histogram). The percentage of exoplanets that is detectable with our space-based MIR interferometer at quadrature (in at least one of the three simulated filters) in addition to radial velocity is shown by the green histogram. The red vertical line indicates our baseline radial velocity detection threshold of $K_* \geq 10~\text{cm}~\text{s}^{-1}$.}
\label{fig06}
\end{figure}

Regarding the observing strategy, another aspect is that for our Monte Carlo simulations we assumed that all exoplanets are observed at a randomly chosen instant of time by placing the exoplanets at a random position on a randomly oriented orbit. Similar to \citet{quanz2015} we could assume that radial velocity surveys would discover nearby exoplanets in advance and put constraints on their orbits. These constraints could then be used to directly image the exoplanets at quadrature when the apparent angular separation between exoplanet and host star is largest. If we observed all exoplanets at quadrature the expected exoplanet yield would increase by $\sim8\%$ from $\sim315$ to $\sim341$ in our baseline scenario. Figure~\ref{fig06} reveals that a significant percentage of exoplanets could not only be detected by direct imaging with our space-based MIR interferometer, but also by radial velocity if future spectrographs reach a precision of $\lessapprox 1~\text{m}~\text{s}^{-1}$.


Finally, in our baseline scenario we chose a very straightforward observing strategy by looking at each star for the same amount of time. However, by implementing more sophisticated algorithms to optimize the order of targets and observation time per target \citep[e.g., the altruistic yield optimization;][]{stark2014} the expected exoplanet yield could be further increased. Also, one could allow for and optimize revisits \citep[e.g.][]{stark2015} because in some cases it may be more useful to image promising targets multiple times to reach a higher completeness instead of searching for planets around stars from which one would only expect a poor planet yield (cf. Figure~\ref{fig03}).


\subsection{Stellar sample and stellar properties}

Recently, \cite{dressing2017} presented stellar parameters for low-mass dwarfs identified as candidate planet hosts during the \emph{Kepler K2} mission. We compare their derived parameters for stars of spectral types M1V -- M4V to those from our stellar sample to verify the properties we used for the M dwarfs in our simulations. We find that the mean value of our stellar effective temperatures $T_{\text{eff},~*}$ deviates at most by $-7.65\%$ for M1V stars and at least by $-0.30\%$ for M2V stars. Moreover, our stellar radii $R_*$ are significantly underestimated for M1V (mean value $-18.97\%$) and M4V ($-23.01\%$) stars but slightly overestimated for M2V ($+12.07\%$) and M3V ($+4.00\%$) stars. Consequently, on average, we underestimated the luminosity of the M-type stars from our stellar sample, which translates into underestimated planet equilibrium temperatures and therefore underestimated thermal blackbody emission from the exoplanets. Hence, our simulations set a lower limit for the expected number of detectable exoplanets around M dwarfs.


\subsection{Exo-Earths and spectroscopic characterization}
\label{exo_earths}

In our baseline scenario we would detect $\sim9$ Earth twins; which are here defined as planets with $0.5~R_\oplus \leq R_\text{p} \leq 1.25~R_\oplus$ and $200~\text{K} \leq T_\text{eq,~p} \leq 300~\text{K}$. This translates into a $99.99\%$, $95.14\%$, $43.89\%$ chance to find at least $1$, $5$, $10$ Earth twin(s) around stars from our stellar sample. Our planet population, however, does neither contain Earth-like planets orbiting G-type stars with orbital periods $> 300~\text{d}$ nor planets with orbital periods $> 418~\text{d}$ orbiting A- and F-type stars (see Section~\ref{planet_population}). Hence, the number of detectable Earth twins must be seen as a lower limit. Furthermore, as indicated in Section~\ref{observing_strategy}, a higher number can be achieved by longer observation times per target and by further optimizing the observing strategy. For instance, if we had 10  times longer on-source time per target in the detection phase compared to our baseline scenario, the total number of detectable exoplanets would increase from $\sim315$ to $\sim486$ (i.e., an increase by roughly $50\%$). The additional planets that could be discovered would primarily be cool ($T_\text{eq,~p} \leq 400~\text{K}$) and therefore faint. More importantly, the expected number of Earth twins detectable in at least one of the three filters simulated in this work would increase from $\sim9$ in our baseline scenario to $\sim48$. This again emphasizes the need for an optimized observing strategy that takes into account flexible integration times per target star, but also a more realistic noise model.

However, if looking for habitable (or even inhabited) planets is one of the major science goals, then one should keep in mind that the parameter space of interesting planets might actually be larger. Life may exist even under significantly higher temperatures of up to $\sim450~\text{K}$ and one might even find biosignatures in the MIR in $\text{H}_2$-dominated atmospheres of super-Earths with radii of up to $1.75~R_\text{Earth}$ \citep[e.g.,][]{seager2013}. Hence, if we add up all detectable exoplanets with equilibrium temperatures between $200$ and $450~\text{K}$\footnote{We acknowledge that the surface temperatures of planets can be significantly higher than their equilibrium temperatures and hence planets at the high temperature end of this range may be too hot for life to exist.} and radii between $0.5$ and $1.75~R_\text{Earth}$ even in our baseline scenario, we end up with $\sim85$ prime targets for in-depth characterization.

As discussed in Section~\ref{observing_strategy}, in our baseline scenario we spend at most three years in the detection phase so that there would be at least two years left for spectroscopic observations in the characterization phase assuming a nominal mission duration of five years. This would translate into an average time of $6~\text{d}$, accounting for $\sim40\%$ overheads again, available for follow-up observations of each of the $85$ prime targets. This seems to be sufficient for a robust characterization of a few dozens of prime targets, but  a more detailed analysis concerning the required S/N, spectral resolution, and wavelength range is also necessary\footnote{\citet{cockell2009} suggest a wavelength range from $6$ -- $20~\text{\textmu m}$ and a spectral resolution of at least $25$ (possibly $300$).}.


\subsection{Mid-infrared interferometer versus optical/near-infrared telescope}

While it seems clear that the atmospheric characterization of a large sample of (small) nearby exoplanets is one of the long-term goals of exoplanet science, a space-based MIR interferometer is just one possible mission concept. However, even the next generation of ground-based ELTs  only enable the detection of a dozen small to mid-sized exoplanets around the most nearby stars \citep[for a quantitative analysis see, e.g.,][]{crossfield2013, quanz2015}. One interesting and competitive alternative would be a large aperture, space-based optical/NIR telescope optimized for high-contrast exoplanet imaging observations in reflected host star light. Concepts for such a mission are currently studied by NASA (e.g., \emph{HabEx} \citep{mennesson2016} and \emph{LUVOIR} \citep{peterson2017}).

\begin{table}[h!]
\caption{\label{habex_baseline} Instrumental parameters for our baseline scenario for \emph{HabEx}/\emph{LUVOIR}.}
\centering
\begin{tabular}{lll}
\hline\hline
\noalign{\smallskip}
Parameter               & Value                           & Description \\
\noalign{\smallskip}
\hline
\noalign{\smallskip}
$D$                     & $12~\text{m}$                   & Aperture size \\
IWA                     & $2~\lambda_\text{eff}/D$        & Inner working angle \\
$C_\text{ref}$          & $1\mathrm{e}{-10}$              & Achievable contrast performance \\
$\lambda_\text{cen,~V}$ & $554~\text{nm}$                 & Central wavelength of V-band filter \\
$\lambda_\text{cen,~J}$ & $1245~\text{nm}$                & Central wavelength of J-band filter \\
$\lambda_\text{cen,~H}$ & $1625~\text{nm}$                & Central wavelength of H-band filter \\
$F_\text{lim,~V}$       & $3.31\mathrm{e}{-10}~\text{Jy}$ & Sensitivity limit (V-band)\tablefootmark{a} \\
$F_\text{lim,~J}$       & $9.12\mathrm{e}{-10}~\text{Jy}$ & Sensitivity limit (J-band)\tablefootmark{a} \\
$F_\text{lim,~H}$       & $8.32\mathrm{e}{-10}~\text{Jy}$ & Sensitivity limit (H-band)\tablefootmark{a} \\
\noalign{\smallskip}
\hline
\end{tabular}
\tablefoot{\\
\tablefoottext{a}{from \url{http://jt-astro.science:5101/hdi_etc} (as of 17 July 2017) assuming an exposure time of $9.7~\text{h}$, a S/N of $\sim10$ and a filter zero point in the AB system of $3631~\text{Jy}$}
}
\end{table}


\begin{figure}[h!]
\centering
\includegraphics[width=\hsize]{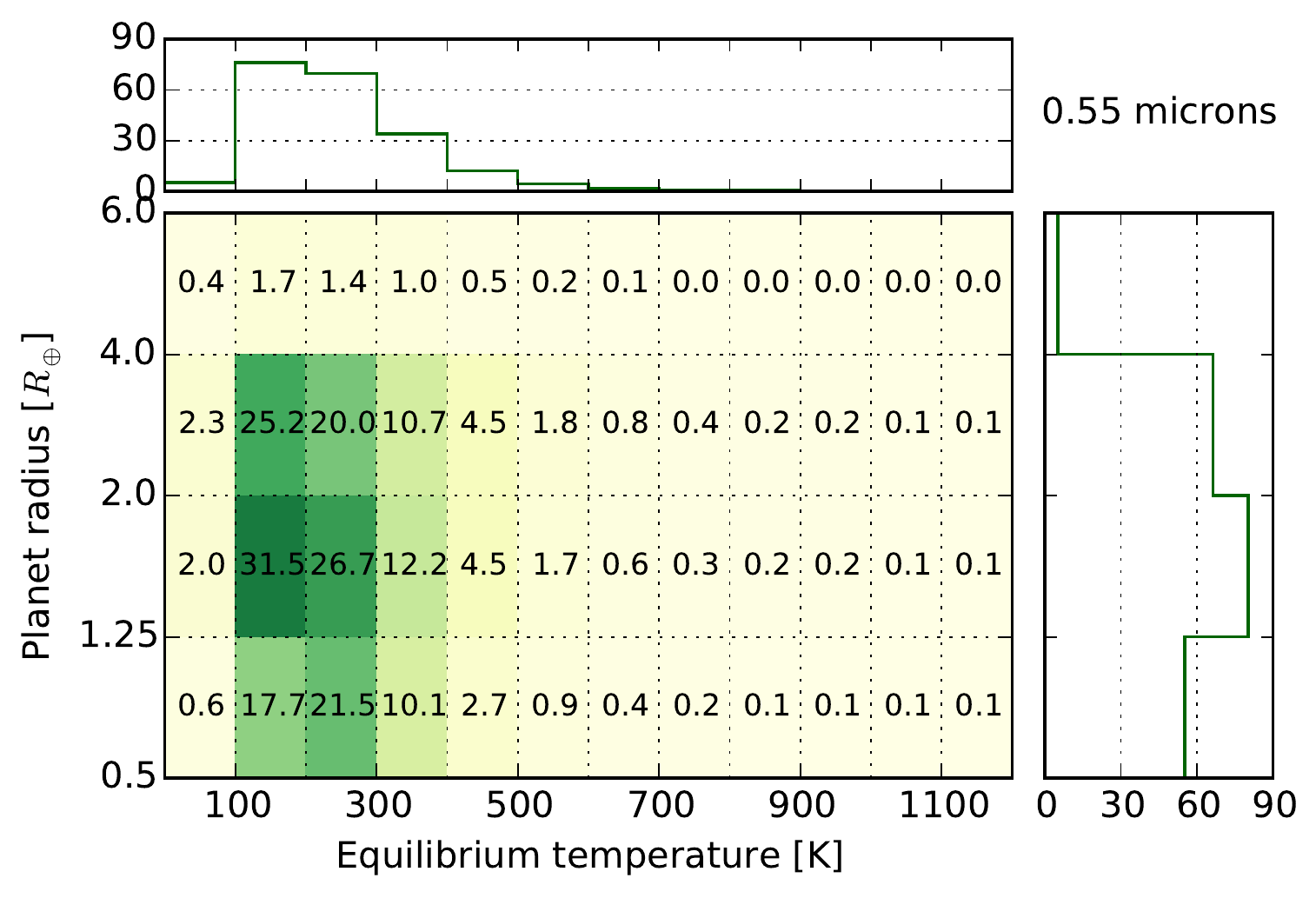}
\includegraphics[width=\hsize]{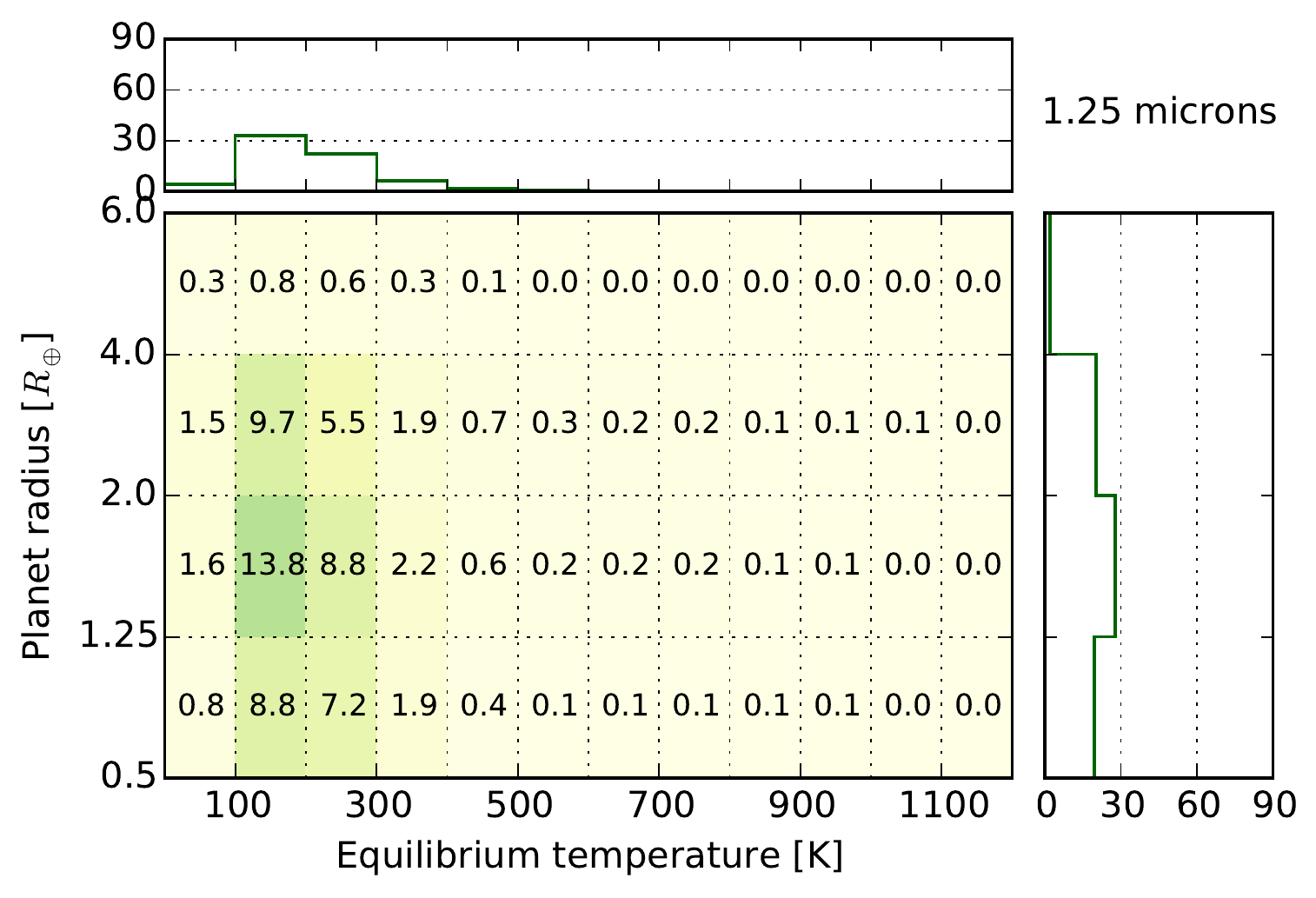}
\includegraphics[width=\hsize]{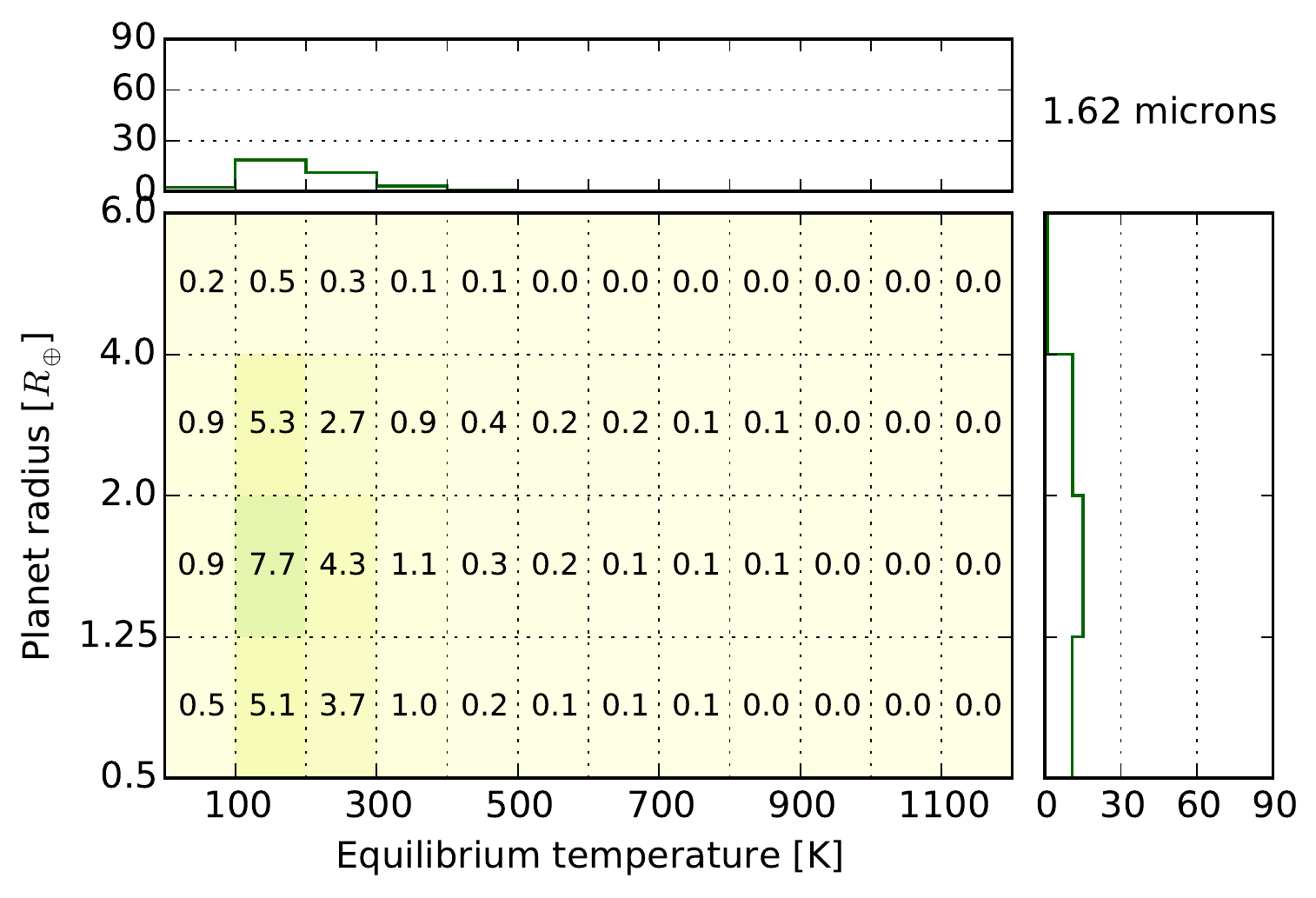}
\caption{Expected number of detectable exoplanets binned in the radius-equilibrium temperature plane (like in Figure~\ref{fig01}) for a large aperture, space-based optical/NIR telescope assuming the instrument parameters presented in Table~\ref{habex_baseline} observing in the V band (top), J band (middle) and H band (bottom). The geometric albedos of our planet population are now distributed uniformly between $0$ and $0.6$ and the x-axis is shifted by $100~\text{K}$ toward cooler equilibrium temperatures.}
\label{fig07}
\end{figure}


\begin{figure}[h!]
\centering
\includegraphics[width=\hsize]{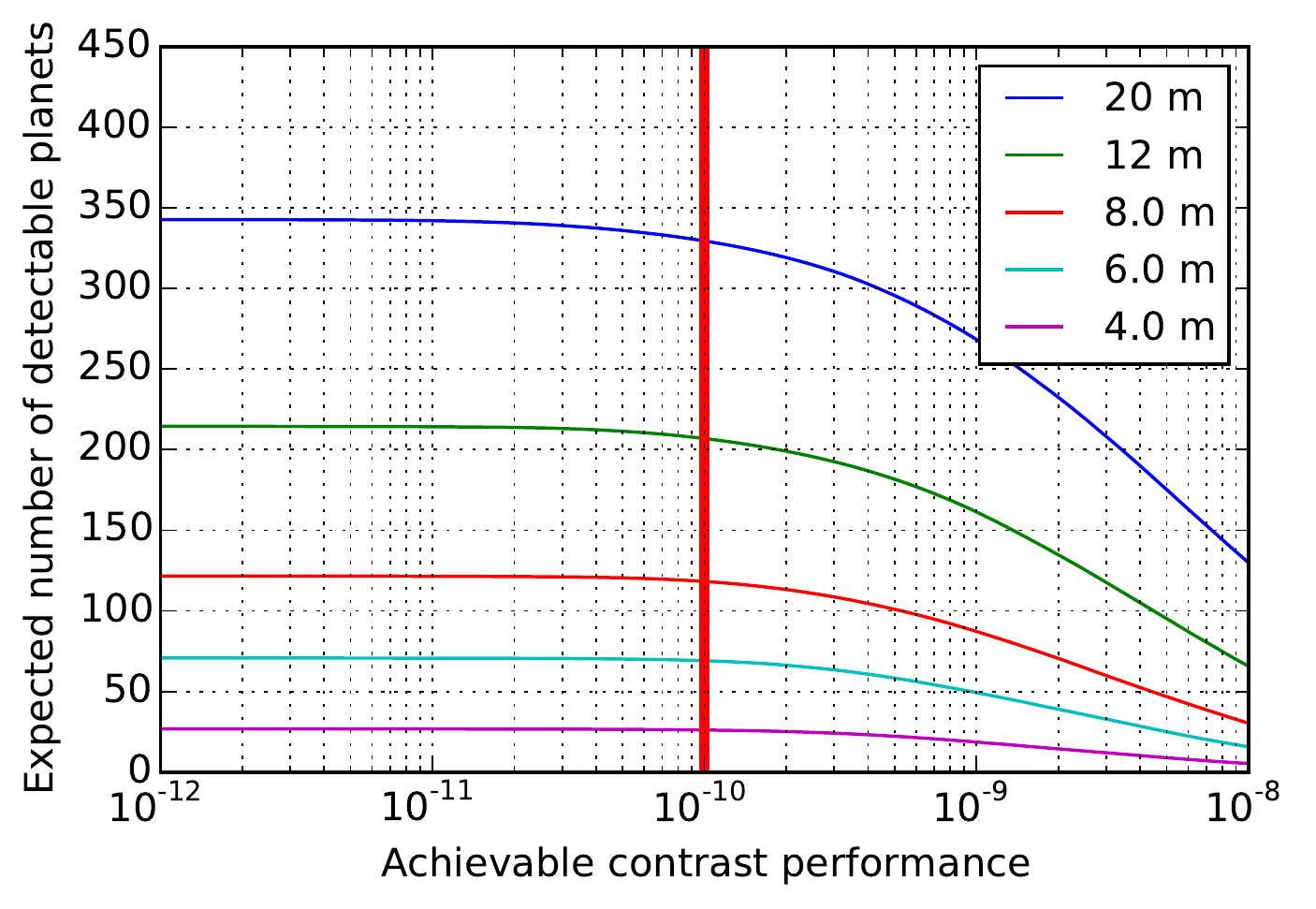}
\caption{Expected number of detectable exoplanets (with radii $< 6~R_\text{Earth}$) as a function of the achievable contrast performance $C_\text{ref}$ in the V band. The five curves represent different aperture diameters $D$ from $20~\text{m}$ to $4.0~\text{m}$. We adapt the sensitivity limit $F_\text{lim,~V}$ according to the aperture size $D$. The red vertical line indicates our baseline contrast performance of $C_\text{ref} = 1\mathrm{e}{-10}$.}
\label{fig08}
\end{figure}


In order to compare, to first order, the expected exoplanet yield of both approaches we also quantified the scientific yield of \emph{HabEx}/\emph{LUVOIR} using our Monte Carlo simulations with the same underlying planet population. The instrument parameters, which we assumed for our baseline scenario, are presented in Table~\ref{habex_baseline}. However, since we are now interested in host star light reflected by the exoplanets, the choice of the geometric albedos had a large impact on our results. Considering the optical/NIR albedos of the solar system planets \citep[cf., e.g.,][]{seager2010} we distributed new geometric albedos uniformly between $0$ and $0.6$ to our exoplanet population. We did not assume any OWA, but we adapted a sensitivity limit corresponding to the $10\sigma$ detection limit in $35'000~\text{s}$ observing time for the MIR interferometer (see Table~\ref{habex_baseline}). For a fair comparison, we also ignored the impact of (exo-)zodiacal light. A planet is considered detectable if its apparent angular separation is larger than the IWA of the instrument, its flux contrast relative to the host star is larger than $C_\text{ref}$, and its total flux observed in a particular band exceeds $F_\text{lim}$ (cf. Table~\ref{habex_baseline}).

The expected numbers of detectable exoplanets for \emph{HabEx}/\emph{LUVOIR} are shown in Figure~\ref{fig07}. In our baseline scenario $\sim207$ planets could be detected in the V band, but only $\sim70$ and $\sim38$ in the J band and  H band, respectively. This has a significant impact on the potential to characterize the detectable exoplanets over a broader wavelength range. However, while the overall number of detectable exoplanets is smaller, there are more planets detected at lower equilibrium temperatures compared to our space-based MIR interferometer. In particular there are $\sim22$ Earth twins, which can be detected in the V band compared to the $\sim9$ detectable Earth twins in the baseline scenario of our MIR interferometer. However, looking again at the expected number of detectable exoplanets with equilibrium temperatures between $200$ and $450~\text{K}$ and radii between $0.5$ and $1.75~R_\text{Earth}$, \emph{HabEx}/\emph{LUVOIR} ($\sim63$ detections) performs worse than our MIR interferometer ($\sim85$ detections).

One of the most important findings, however, is the impact of the contrast performance. Figure~\ref{fig08} shows the expected exoplanet yield in dependence of the achievable contrast performance for different aperture diameters; this figure reveals that the expected number of detectable exoplanets with a $12~\text{m}$ aperture does not increase significantly if instead of $1\mathrm{e}{-10}$ we assume $1\mathrm{e}{-11}$ ($\sim214$ detections in the V band) or even $1\mathrm{e}{-12}$ (also $\sim214$ detections in the V band). Lowering the contrast performance, however, has a large impact. In case of $C_\text{ref} = 1\mathrm{e}{-9}$ or $C_\text{ref} = 1\mathrm{e}{-8}$ the expected number of detectable exoplanets in the V band would decrease to $\sim161$ and $\sim66$, respectively. This reveals that a contrast performance of $1\mathrm{e}{-10}$ is close to the optimal regime and further gains in planet yield could only be achieved by decreasing the IWA, i.e., increasing the aperture. Enhancing the sensitivity does not have the same effect as for the MIR interferometer as even in the case of no sensitivity limit at all the expected number of detectable exoplanets in the V band is only $\sim234$.

Looking at the split of detectable exoplanets by host star spectral type shows that \emph{HabEx}/\emph{LUVOIR} has a similar preference for planets around M-type stars, which also make up $\sim50\%$ of all detectable exoplanets.


\section{Summary and conclusions}

We carry out Monte Carlo simulations predicting, for the first time, the expected number of directly detectable (small) exoplanets for a space-based MIR nulling interferometer. On the basis of planet occurrence statistics from the NASA \emph{Kepler} mission, an input catalog of $326$ host stars within $20~\text{pc}$, and the technical specifications from the original mission proposal of the \emph{Darwin} mission \citep{cockell2009}, we find a total expected exoplanet yield of $315_{-77}^{+113}$ (only counting planets with radii $0.5~R_\text{Earth} \leq R_\text{p} \leq 6~R_\text{Earth}$), where the uncertainties are dominated by statistical errors in the underlying planet population. Our baseline scenario assumes an initial detection phase observing in three bands ($5.6$, $10$ and $15~\text{\textmu m}$); roughly $244$ exoplanets are detected in at least two bands. Slightly less than half of all detectable planets orbit low-mass M-type stars, whereas the other half orbit F-, G-, and K-type stars.

By optimizing the observing strategy and assuming a nominal mission duration of five years such a mission should allow for the spectroscopic characterization of a few dozens of habitable or even inhabited planets with effective temperatures between $200$ and $450~\text{K}$ and radii between $0.5$ and $1.75~R_\text{Earth}$. More quantitative investigations regarding the required S/N, spectral resolution, and wavelength range for such a characterization phase are foreseen for a future publication.

The huge sample of directly detectable exoplanets would offer unprecedented opportunities for exoplanet science. The atmospheric composition of planets covering a large range of effective temperatures, radii, and host star spectral types would be a unique dataset for planet formation and evolution studies. This is particulary the case when combined with empirical mass estimates or constraints, for example, from radial velocity, which seems feasible for a significant percentage of the detectable exoplanets. For a considerable subset of these planets the MIR spectra are probed for indications of biosignatures in their atmospheres, addressing the fundamental question of exoplanet habitability. Comparing our MIR nulling interferometer with an optical/NIR telescope further shows that both approaches are extremely promising in themselves and in fact only data from both missions combined could yield planetary radii and albedos for a comprehensive sample of exoplanets.

While finding habitable or even inhabited planets is one of the major objectives of exoplanets research, it seems that only a large/flagship space mission is able to characterize a considerable sample of exoplanets in the most interesting and promising size and temperature range. Even the next generation of $30$ to $40~\text{m}$ ground-based telescopes or \emph{WFIRST-AFTA} will likely only detect a handful, maybe a dozen of objects \citep[e.g.,][]{crossfield2013, quanz2015}. While these projects are a tremendous challenge, they are an absolutely necessary and ground-breaking step for a variety of astrophysical research topics, but in the long run the exoplanet community will have to push for a large, and maybe even dedicated, space mission, to address the (most) relevant questions properly. As demonstrated above, a space-based MIR interferometer mission seems to be, as expected, a very promising concept and with our Monte Carlo tool at hand we have the possibility to quantify the scientific yield of certain design choices as we continue to tackle the technical challenges. The next key steps in this direction will be, first, implementing a more realistic noise model including the effects of exozodical light and stellar leakage and, second, looking in more detail into null-depth, stability and throughput performance of different array configurations \citep[cf.][]{guyon2013}.

In addition, because we know the targets we want to look at, we should start characterizing these stellar systems with available instruments. This includes (a) robust stellar parameters such as metallicity, rotation period, and rotation axis, (b) the search for (massive) planets and unknown substellar companions via radial velocity and high-contrast imaging, and (c) the derivation of at least constraints on the level of exozodiacal dust. Parts of these investigations are already ongoing or will, to some extent, be addressed in the near future. However, if a space-based direct imaging mission is the goal, better coordinated and systematic efforts are required.

The findings from \emph{Kepler} have taught us that moving forward with \emph{Darwin} would not have been in vain. Now it is time to reinitiate the discussion about the next big steps. The lead times are long and in case of Europe the first three slots for large missions within the ESA Cosmic Vision program are already taken by other research fields\footnote{\url{http://sci.esa.int/cosmic-vision/42369-l-class-timeline/}}. However, this is also an opportunity and a coordinated attempt from within the exoplanet community speaking with one voice would be a viable way forward.


\begin{acknowledgements}
This work has been carried out within the frame of the National Center for Competence in Research PlanetS supported by the Swiss National Science Foundation. SPQ acknowledges the financial support of the SNSF. This research has made use of the SIMBAD database, operated at CDS, Strasbourg, France, and of data products from the Two Micron All Sky Survey, which is a joint project of the University of Massachusetts and the Infrared Processing and Analysis Center/California Institute of Technology, funded by the National Aeronautics and Space Administration and the National Science Foundation. The authors gladly thank Ian Crossfield for sharing the stellar catalog, Denis Defr\`ere and Olivier Absil for helpful comments, and the referee for a timely and constructive report that helped to improve the original manuscript.
\end{acknowledgements}

%
%

\bibliographystyle{aa}
\bibliography{v13_jk_final}

\end{document}